\documentclass[prl,noshowpacs,english,superscriptaddress,twocolumn]{revtex4-1}
\usepackage[colorlinks,linkcolor=blue,anchorcolor=blue,citecolor=blue,filecolor=blue,menucolor=blue,runcolor=blue,urlcolor=blue,frenchlinks=blue]{hyperref}
\usepackage{amsmath,amssymb,amsthm,amsfonts,mathrsfs}
\usepackage{graphicx}
\usepackage{pdfpages}
\usepackage{appendix}
\usepackage{lipsum}
\usepackage{slashed}
\usepackage{dcolumn}
\usepackage{mathrsfs}
\usepackage{bm}
\usepackage{color}
\usepackage{lineno,hyperref}
\makeatletter

\newcommand{\Rmnum}[1]{\expandafter\@slowromancap\romannumeral #1@}
\AtBeginDocument{\let\LS@rot\@undefined}
\makeatother
\usepackage{tikz,pgf}
\usetikzlibrary{shadings}
\usepackage{pgfplots,pgfplotstable}

\usepackage{tikz,pgf}
\usetikzlibrary{shadings}
\usepackage{pgfplots,pgfplotstable}

\def \H {\mathcal{H}}

\def \k {\mathbf{k}}

\def \K {\mathcal{K}}

\def \T {\mathcal{T}}
\def \C {\mathcal{C}}
\def \S {\mathcal{S}}

\def \TT {\hat{\T}}
\def \CC {\hat{\C}}
\def \SS {\hat{\S}}
\def \M {\mathcal{M}}
\def \MM {\hat{\M}}

\def \R {\mathcal{R}}
\def \RR {\hat{\R}}

\def \I {\mathcal{I}}
\def \II {\hat{\mathcal{I}}}

\def \Z {\mathbb{Z}}
\def \T {\mathcal{T}}
\def \RT {\mathcal{RT}}

\begin{document}

	\title{Topological Insulators with Hybrid-order Boundary States}

	\author{Yan-Qing Zhu}
	\email{zhuyq1992@gmail.com}
	\affiliation{Quantum Science Center of Guangdong-Hong Kong-Macao Greater Bay Area, 3 Binlang Road, Shenzhen, China}
	\affiliation{Guangdong-Hong Kong Joint Laboratory of Quantum Matter, Department of Physics, \\and HK Institute of Quantum Science $\&$ Technology,\\ The University of Hong Kong, Pokfulam Road, Hong Kong, China}
	\affiliation{Key Laboratory of Atomic and Subatomic Structure and Quantum Control (Ministry of Education), Guangdong Basic Research Center of Excellence for Structure and Fundamental Interactions of Matter, School of Physics, South China Normal University, Guangzhou 510006, China}

	\author{Zhen Zheng}
	\affiliation{Key Laboratory of Atomic and Subatomic Structure and Quantum Control (Ministry of Education), Guangdong Basic Research Center of Excellence for Structure and Fundamental Interactions of Matter, School of Physics, South China Normal University, Guangzhou 510006, China}
\affiliation{Guangdong Provincial Key Laboratory of Quantum Engineering and Quantum Materials, Guangdong-Hong Kong Joint Laboratory of Quantum Matter, Frontier Research Institute for Physics, South China Normal University, Guangzhou 510006, China}

	\author{Giandomenico Palumbo}
	\email{giandomenico.palumbo@gmail.com }
	\affiliation{School of Theoretical Physics, Dublin Institute for Advanced Studies,
		10 Burlington Road, Dublin 4, Ireland}
	
	\author{Z. D. Wang}
	\email{zwang@hku.hk}
	\affiliation{Guangdong-Hong Kong Joint Laboratory of Quantum Matter, Department of Physics, \\and HK Institute of Quantum Science $\&$ Technology,\\ The University of Hong Kong, Pokfulam Road, Hong Kong, China}
	\affiliation{Quantum Science Center of Guangdong-Hong Kong-Macao Greater Bay Area, 3 Binlang Road, Shenzhen, China}
	\affiliation{Guangdong Provincial Key Laboratory of Quantum Engineering and Quantum Materials, Guangdong-Hong Kong Joint Laboratory of Quantum Matter, Frontier Research Institute for Physics, South China Normal University, Guangzhou 510006, China}

	\date{\today}
	
	\begin{abstract}
		We report the discovery of several classes of novel topological insulators (TIs) with hybrid-order boundary states generated from the first-order TIs with additional crystalline symmetries. Unlike the current studies on hybrid-order TIs where different-order topology arises from merging different-order TIs in various energy, {these novel TIs exhibit unique properties, featuring a remarkable coexistence of first-order gapless modes and higher-order Fermi arc states}, behaving as a hybrid between the first-order TIs and higher-order topological semimetals within a single bulk gap.  Our findings establish a profound  connection between these  novel $d$-dimensional ($d$D) TIs and ($d-1$)D higher-order TIs (HOTIs), which can be understood as a result of stacking $(d-1)$D HOTIs to $d$D with $d=3,4$, revealing unconventional topological phase transitions by closing the gap in certain first-order boundaries rather than the bulk. 
		The bulk-boundary correspondence between these higher-order Fermi-arcs and  bulk topological invariants acossiated with additional crystallline symmetries is also demonstrated.
		We then adress the conventional topological phase transitions from these novel TIs to nodal-line/nodal-surface semimetal phases, where the gapless phases host new kinds of topological responses. 
		Meawhile, we present the corresponding tolopogical semimetal phases by stacking these unique TIs.
	  Finally, we discuss potential ways to realize these novel phases in synthetic and real materials, with a particular focus on the feasible implementation in optical lattices using ultracold atoms.

	\end{abstract}
	
	\maketitle

	\emph{\color{blue}Introduction.---}
Topological insulators (TIs) and topological semimetals (TSMs) have emerged as one of the most active fields of modern physics over the past two decades \cite{Hasan2010,XLQi2011,Armitage2018}, attracting significant attention from researchers in condensed matter physics and artificial systems\cite{DWZhang2018,Cooper2019,YXu2019,Ozawa2019b,LLu2014,JLiu2020,WZhu2023}. One significant property of topological phases is the presence of bulk-boundary correspondence, which guarantees the existence of gapless first-order boundary modes that are associated with certain bulk topological invariant.  These first-order topological materials have been classifed within the framework provided by the real K-theory \cite{Atiyah1966} and Altland-Zirnbauer (AZ) classes \cite{CKChiu2016} for both gapped \cite{Schnyder2008,Kitaev2009} and gapless \cite{Horava2005,YXZhao2013,YXZhao2014} systems, based on their fundamental symmetries including time-reversal $\T$, charge-conjugation $\C$, and chiral  symmetry $\S$.

Recent research focus has been extended to higher-order topological phases with additional crystalline symmetries \cite{Benalcazar2017,ZDSong2017,Langbehn2017,Schindler2018a,Schindler2018,LMao2018,Roy2019,Roy2020,ZWang2019,Wieder2020,YYang2023}.  These systems do not feature first-order even second-order boundary gapless states but instead exhibit gapless states at higher-order boundaries. 
For instance, a second-order TI (SOTI) in 3D hosts boundary gapless modes at 1D hinges (i.e., second-order boundary) while its 2D surface (i.e., first-order boundary) spectrum is gapped. By further gapping the gapless hinge states,  a 3D third-order TI (TOTI) can be induced, leading to the existence of zero-energy modes at 0D corners (i.e., third-order boundary).  By stacking these HOTIs along an extra dimension, one can obtain higher-order TSMs (HOTSMs) with higher-order Fermi-arcs \cite{LMao2018,ZWang2019,Wieder2020,HXWang2020,Ghorashi2020,Simon2022,QWei2021a,Rui2022,YYang2023}.
Moreover, hybrid-order TIs \cite{Kooi2020} with multi-gaps have been also discovered in metamaterials \cite{XZhang2020,YYang2021,HSLai2023}, which simultaneously host hybrid-order topological boundary modes  under different open boundary conditions (OBCs) in various energy. Building on recent advancements in the aforementioned areas, a natural question arises: \emph{Are there novel classes of TIs that host hybrid-order boundary modes beyond the above regimes? Can we establish bulk-boundary correspondence in these phases if they exist?}
  
In this Letter, we answer these questions positively. We introduce several novel classes of TIs that exhibit unique hybrid-order boundary states consisting of both first-order gapless modes and higher-order Fermi-arc states within a single bulk gap, which can be seen as a hybrid between first-order TIs (FOTIs) and HOTSMs. These phases can be simply generated from the first-order TIs (FOTIs) by shifting the gapless modes in first-order boundary Brillouin Zone (BZ) through specific perturbations protected by crystalline symmetries. On the other hand, we can also regard
these unique  TI phases in $d$D as the stacking $(d-1)$D HOTIs along an additional spatial dimension ($d=3,4$).  In this viewpoint, we reveal unconventional topological phase transitions (UTPTs) by closing the first-order boundary gap while maintaining an open bulk gap, i.e., the unconventional phase transition points are topological crystalline insulators (TCIs) \cite{LFu2011,Slager2013}. We emphasize these higher-order Fermi-arcs can be well described by the bulk topology associated with related crystalline symmetries. {All the definitions of the relevant crytalline symmetries are provided in the supplemental material (SM)\cite{SM}.}
For concreteness, we mainly focus on several 4D models in the main text: (i) Class A with combined mirror symmetry; (ii) Class AIII with $\RT$-symmetry ; (iii) Class A with $C_4\T$-symmetry; (iv) Class AIII with $C_4\T$-symmetry. {The corresponding topological properties of these phases, including a summary table, are detailed in the SM.}
In (i), we find a 4D TI hosts both first-order boundary Weyl cones and second-order chiral hinge Fermi-arcs located at two hinges; In (ii), a 4D TI with spin conservation hosts first-order boundary Dirac cones and second-order helical hinge Fermi-arcs while Dirac cones will be expanded into real Dirac nodal-lines and helical hinge states will be gapped leading to third-order Fermi-arcs after introducing a spin-orbit-coupling (SOC) $\RT$-symmetry perturbation; In (iii), a 4D TI with $C_4\T$ is similar to the case (i) but with the difference that the second-order hinge Fermi-arc are always located at four hinges; (iv) Two copies of Hamiltonian with opposite sign in (iii) are considered which supports first-order boundary Dirac cones and second-order helical hinge Fermi-arcs.  
By performing dimensional reduction with $k_z=0$, the corresponding 3D models and physical pictures are obtained, which are presented in the SM.  The explicit expressions of the topological numbers associated with the related crystalline symmetries are presented. Moreover, we also explore the conventional topological phase transition (CTPT) from a novel TI in case (i)/(ii) to a nodal-line/nodal-sphere phase, presenting the topological response theory of these gapless phases. 
Furthermore, we discuss the related topological semimetals by stacking these novel TIs as usual. Finally, we discuss the possible ways to realize these novel phases in synthetic and real materials, including the implemented proposals in optical lattices using ultracold atoms.

\begin{figure}[http]\centering
		\includegraphics[width=8cm]{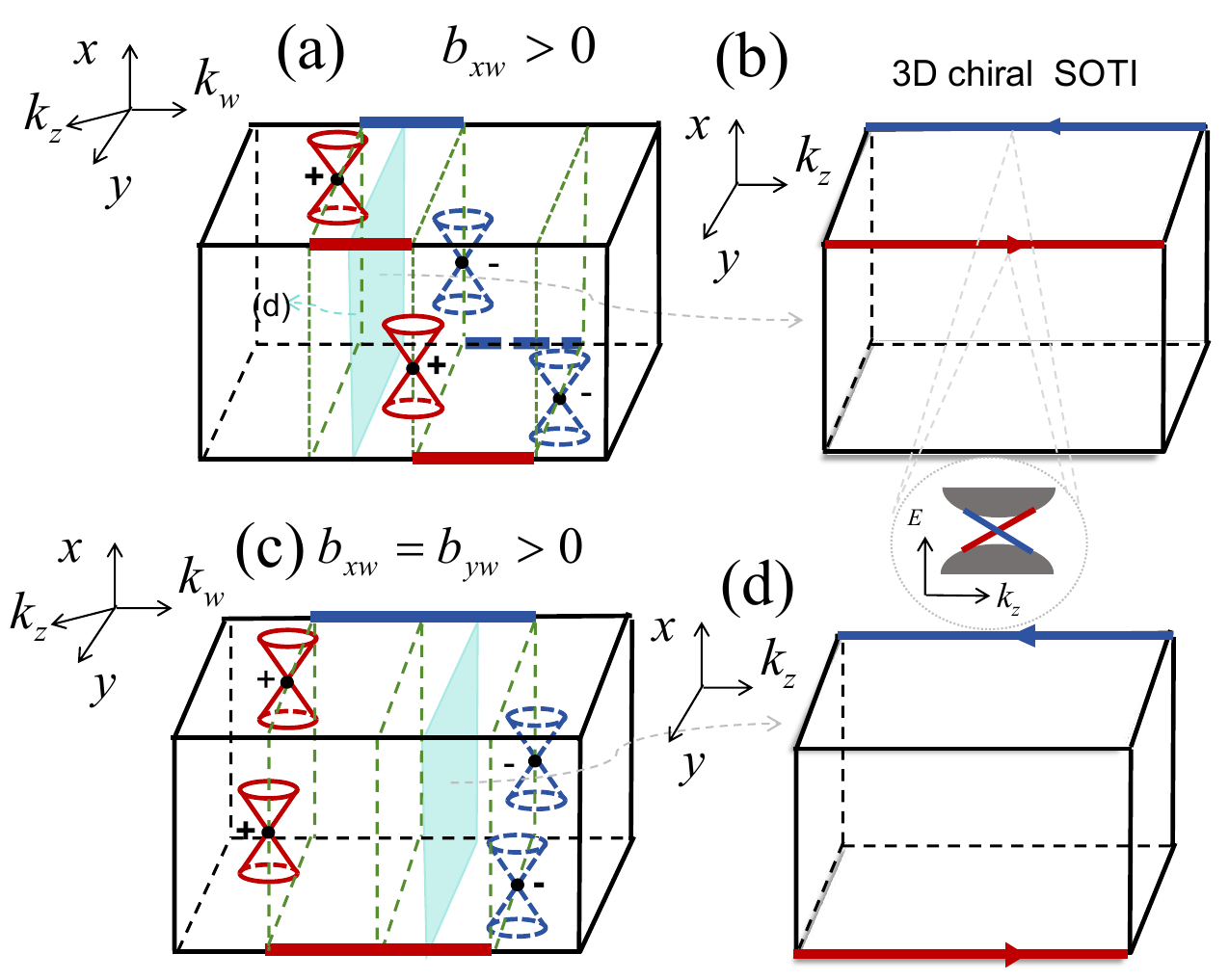}
		\caption{ Schematic of the boundary states of  $\H_1$ with $C_2=-1$ when $2<M<4$. Each first-order 3D boundary hosts a 3D Weyl cone marked in red/blue carrying right-/left- chirality while the 1D chiral hinge modes are also marked in the same way. (a) $b_{xw}$ term  shifts the Weyl cone along the $k_w$ direction, causing it to be located at $\k_{x^+}=(k_y,k_z,k_w)=(0,0,b_{xw})$ at $x=-L/2$, while the Weyl cone at $x=L/2$ is shifted and located at $\k_{x^-}=(0,0,-b_{xw})$. If we further open the boundary along the $y$ axis, there will be 1D chiral zero-energy Fermi-arcs located at the symmetry-protected hinges for $k_z=0$. { (b) For $k_{w}\in (-b_{xw},0)$, 3D subsystems act as  chiral SOTIs that host the chiral hinge modes located the hinge positions at $(x,y)=(L/2,\pm L/2)$ when considering OBC along $x$ and $y$ direcitions.
		(c) The case when $b_{xw}=b_{yz}$ where second-order Fermi-arcs only appear at two $\M_{(x-y)z}$-symmetric hinges.
		(d) 3D subsystems $\H^{k_w}_1$ act as chiral SOTI in the region $k_w\in (-b_{xw},0)\cup (0,b_{xw})$, hosting chiral hinge modes located at the hinge positions $(x,y)=(\mp L/2, \pm L/2)$ with mirror reflection plane about the $x=y$ plane. The inset figure depicts the chiral modes as 
        energy-momentum $E$-$k_z$ at each hinge, highlighting the different chiralities with red and blue colors.
		}}
		\label{Fermiarc}
	\end{figure}

	\emph{\color{blue}Four-band model in Class A$^{\M_{ij}}$.---} We now consider a 4D four-band model with its Bloch Hamiltonian given by,
	\begin{equation}
		\begin{split}\label{4Dmodel}
			&\H_1(\boldsymbol k)=\H_0+\Delta_{1},\\
		\end{split}
	\end{equation}
where $\H_0=\sum_i d_i\Gamma_i+d_0\Gamma_0$  with the Bloch vector $d_i=\sin k_i$,  $d_0=M-\sum_i \cos k_i$ ($i=x,y,z,w$). The perturbation 
$\Delta_1=b_{xw}\Gamma_0\Gamma_{14}+b_{yw}\Gamma_0\Gamma_{24}$.
Here the Driac matrices are  $\Gamma_1=G_{31}$, $\Gamma_2=G_{32}$, $\Gamma_3=G_{33}$, $\Gamma_4=G_{20}$, $\Gamma_0=G_{10}$, and $\Gamma_{ab}=\frac{i}{2}[\Gamma_a,\Gamma_b]$, satisfying $\{\Gamma_i,\Gamma_j\}=2\delta_{ij}$.  We mix the notation $i\in(1,2,3,4)\leftrightarrow (x,y,z,w)$ in the subscript of $\Gamma_i$ and label $G_{ab}=\sigma_a\otimes\sigma_b$ through the $2\times 2$ identity $\sigma_0$ and Pauli matrices $\sigma_i$ hereafter. As one can see that $\H_0$ is the well-known 4D Chern insulator characterized by the second Chern number (SCN) $C_2$ \cite{SCN},  harboring $|C_2|$ Weyl cones at the origin of the 3D boundary BZ \cite{XLQi2008}. Under OBCs along the $x$ ($w$)-direction with a chain length $L$, introducing a non-zero $b_{xw}$ term in $\H_0$ leads to the shifted Weyl cones along the $k_w$ ($k_x$)-axis in the boundary BZ, resulting in the second-order chiral Fermi-arc hinge states when considering OBCs at $x$ and $y$ directions 
, as shown in Fig. \ref{Fermiarc} (a). We present the effective boundary Hamiltonian and numeric results in the SM \cite{SM}.

In what follows we discuss the topology of this system. The initial Hamiltonian $\H_0(\boldsymbol k)$ in class AII respects $\T$ represented as $\TT=\Gamma_{13}\K$, where $\K$ is the complex conjugate. In addition,  $\H_0$ preserves extra crystalline symmetries, including inversion symmetry $\I$ \cite{IS} and the combined mirror reflection symmetries $\M_{ij}=\M_i\M_j$, where each mirror reflection $\M_i$ inverts one of the momentum components $k_i$ \cite{MS}.  
For simplicity, we first consider the case when only $b_{xw}$ term in $\Delta_1$ is numerically small, it breaks $\T$ and only preserves $\M_{xw}$ and $\M_{yz}$ symmetries. The system now falls into class A and still hosts the unchanged $C_2$ with the shifted first-order Weyl boundary states.
To characterize the second-order topology, we can define extra topological numbers at $\M_{xw}$/$\M_{yz}$-invariant points $\Lambda_a^{xw}/\Lambda_a^{yz}$ \cite{HSP}, i.e., named ``\emph{mirror first Chern number}" (MFCN), 
\begin{equation}\label{MC}
C_{1m}^{ij}(\Lambda_a^{ij})=[C_{+i}(\Lambda_a^{ij})-C_{-i}(\Lambda_a^{ij})]/2,
\end{equation}
where $C_{\pm i}$ denotes the first Chern number defined in 2D subsystems $\H_1^{\Lambda_a^{ij}}(k_l,k_m)$ of the eigenspace $\mathcal{E}=\pm i$ of $\MM_{ij}$ \cite{MS}.
	For simplicity and without loss of generality, we focus on the case when $M$ near $3$ with $C_2=-1$ hereafter, evaluation shows that $C_{1m}^{xw}=-1$/$C_{1m}^{yz}=1$ at $\Lambda_1^{xw}/\Lambda_1^{yz}=(0,0)$, and $C_{1m}^{xw}(\Lambda_a^{xw})=0$/$C_{1m}^{yz}(\Lambda_a^{yz})=0$ for the rest of the three $\Lambda_a^{xw/yz}$ points \cite{SM}.
These results establish the bulk-hinge correspondence and provide detailed information of the chiral hinge Fermi-arcs, as depicted in Fig. \ref{Fermiarc}(a).
	Since these hinge Fermi-arcs are formed by stacking the 1D chiral modes $\H_{R/L}=\pm k_z$ along $k_w$ axis,  once $\M_{xw}$ acts on one of hinge states, it will transform these states to the mirror reflection hinge about $x=0$ and $k_w=0$ planes without changing its chirality;  while $\M_{yz}$ tranforms one of the hinge states to its mirror-symmetric hinge about $y=0$ and then changes the chirality of this hinge mode due to the mirror reflection about $k_z=0$ plane, causing $\pm k_z\rightarrow \mp k_z$. Similar picture appears when $M$ is in other parameter regions\cite{SM}. {Additionally, a comparable case occurs when only $b_{yw}$ is non-zero.}
	
When both $b_{xw}$ and $b_{yw}$ survive, $\M_{xw}$ and $\M_{yz}$ are broken. For the typical case when $b_{xw}=b_{yw}$, $\H_1$ hosts an extra combined symmetry $\M_{(x-y)z}=\M_{x-y}\M_z$ with $\M_{x-y}$ being the mirror reflection about the mirror-invariant line $k_x=k_y$\cite{N3}.
In the same spirit, we can calculate the MFCNs $C_{1m}^{k_z}$ at $k_x=k_y$, $k_z=0,\pi$ planes with $C_{1m}^{0}=-1$ and $C_{1m}^{\pi}=0$\cite{SM}.
As we can see that chiral Fermi-arcs are located at diagonal hinges with mirror reflection about $x=y$ plane harboring opposite chirality due to the operation of $\M_z$, as shown in Fig. \ref{Fermiarc}(c-d).  
Similarly, when $b_{xw}=-b_{yw}$, $\H_1$ preserves $\M_{(x+y)z}$ symmetry  with $C_{1m}^{0}=1$ and $C_{1m}^{\pi}=0$, enforcing
the location of Fermi-arcs at anti-diagonal corners with opposite chirality\cite{SM}.  If $b_{xw}\neq |b_{yw}|$, $\Delta_1$  breaks $\T$ and all the $\M_{ij}$-symmetries. However, the system still hosts the $\I$-symmetry which guarantees the second-order Fermi-arc states as a superposition of the cases where either $b_{yw}=0$ or $b_{xw}=0$. We emphasize that $\M_{ij}$-symmetry is used to construct this unique TI; it is itself not essential for the existence of second-order hinge Fermi-arcs. The hinge states are robust against an $\M_{ij}$-symmetry breaking perturbation, as long as the bulk gap is not closed.

To understand this 4D novel phase with both first- and second-order boundary states, we can treat  $k_w$ as a parameter  in this $\H_1$.
 The 3D subsystem $\H_1^{k_w}(k_x,k_y,k_z)$ goes through UTPTs while the gap closing happens in certain 2D surface BZs instead of the 3D bulk when tuning $k_w$ from $-\pi$ to $\pi$. For instance and when $b_{xw}>0$, the 3D subsystems exhibit two distinct phases, which are normal insulators for $k_w\in (-\pi, -b_{xw}) \cup(b_{xw},\pi)$, and chiral SOTIs for $k_w\in(-b_{xw},0)\cup(0,b_{xw})$ {[see Fig. \ref{Fermiarc}(b)]}. The TCIs \cite{LFu2011,Slager2013} at $k_w=0,\pm b_{xw}$ serve as the unconventional transition points between these two phases.
  Thus this novel 4D TI phase $\H_1$ is nothing but the stacking layers of 3D chiral SOTIs, revealing UTPTs by tuning $k_w$. We present detailed discussions about the topology of these 3D subsystems  in the SM\cite{SM}.  On the other hand, a 3D novel TI  with hybrid-order boundary states can be simply obtained by setting $k_z=0$, the system restores $\S$-symmetry and thus falls into class AIII. Similar discussions can be found in the SM\cite{SM}.
	

	\emph{\color{blue}Eight-band model in Class AIII$^{\RT}$ without and with SOC.---}
We now consider a 4D eight-band model, its Hamiltonian is given by,
	\begin{equation}
		\begin{split}\label{4Dmodel2}
			\H_2(\boldsymbol k)=\tilde{\H}_0+\Delta_{2}+\Delta_{3},
		\end{split}
	\end{equation}
where $\tilde{\H_0}=\sum_i d_i\tilde{\Gamma}_i+d_0\tilde{\Gamma}_0$  with the Bloch vector $d_i$ $(i=x,y,z,w)$ are the same as in $\H_0$,	the perturbations $\Delta_{2}=b_{xw}\tilde{\Gamma}_{0}\tilde{\Gamma}_{14}+b_{yw}\tilde{\Gamma}_{0}\tilde{\Gamma}_{24}+b_{zw}\tilde{\Gamma}_0\tilde{\Gamma}_{34}$,
and $\Delta_{3}=c_{x}\tilde{\Gamma}_{61}+c_z\tilde{\Gamma}_{63}$.
	Here $\tilde{\Gamma}_1=G_{331}$, $\tilde{\Gamma}_2=G_{332}$, $\tilde{\Gamma}_3=G_{333}$, $\tilde{\Gamma}_4=G_{320}$, $\tilde{\Gamma}_0=G_{310}$, $\tilde{\Gamma}_6=G_{100}$, $\tilde{\Gamma}_7=G_{200}$, and $\tilde{\Gamma}_{ab}=\frac{i}{2}[\tilde{\Gamma}_a,\tilde{\Gamma}_b]$. Label $G_{ijk}=\sigma_i\otimes\sigma_j\otimes \sigma_k$ hereafter. $\tilde{\H}_0$ is just the two copies of 4D Chern insulators $\pm\H_0$ with opposite $\pm C_2$ implying the existence of 3D first-order boundary Dirac modes.
	$\Delta_2$ shifts the gapless Dirac cones in 3D  boundary momentum space similar to that case in $\H_0$, resulting in second-order helical Fermi-arc hinge states. 
 $\Delta_3$ behaves like an SOC, where $c_i$ expands the 3D Dirac cone into a nodal-line  structure at $k_i=0$ plane at the boundary normal to  $j$-direction.  For instance, non-zero $c_x$ term expands the boundary Dirac cone into a noda-line at $k_x=0$ plane in the case of OBC at $y/z/w$-direction\cite{SM}.

In the following we discuss the topology of this model in detail. 
In  AZ classification \cite{CKChiu2016}, $\tilde{\H}_0$ falls into class CII with $\T$, $\C$ and $\S$ hosting a $\Z_2$ topology\cite{N4}. 
 Since $\tilde{\H}_0$ is spin conservation with $\tilde{\Gamma}_5$-symmetry, we have $[\tilde{\Gamma}_5,\tilde{\H}_0]=0$, where $\hat{\tilde{\Gamma}}_5=G_{300}$. We can define a $\Z_2$ number through the spin SCN \cite{N5} to characterize the first-order topology with bulk-boundary correspondence even if $\T$ and $\C$ will be broken later\cite{YQZhu2022}. In addition, $\tilde{\H}_0$ also respects the following crystalline symmetries, including  $\I$, and  $\M_i$ leading to the combined symmetries $\M_{ij}$ and $\R_i\T=\M_i\I\T $ for $i=x,y,z,w$\cite{N6}.

	\begin{figure}[http]\centering
	\includegraphics[width=7cm]{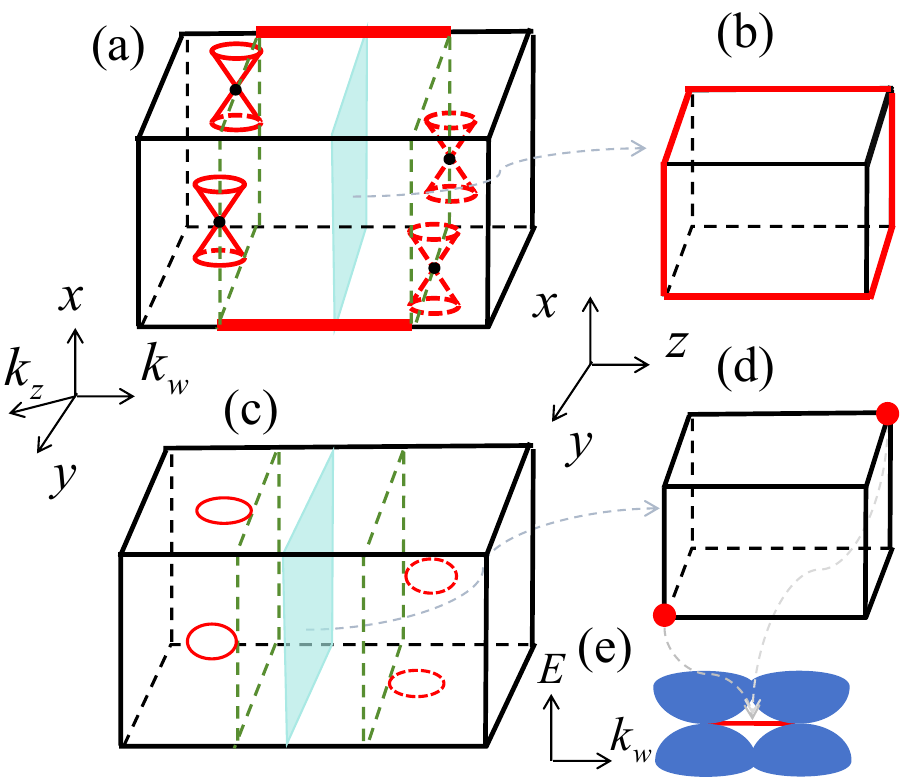}
	\caption{ Schematic of the boundary states of $\H_3$ when $2<M<4$. (a-b)$b_{xw}=b_{yw}=b_{zw}>0$ and $c_x=c_z=0$, the system harbors both shifted first-order Dirac points located at the same positions as in Fig. \ref{Fermiarc}(c) and second-order Fermi-arcs consisting of helical hinge states (colored in orange) of 3D SOTIs stacked along $k_w$ axis. (c-e) $b_{xw}=b_{yw}=b_{zw}>0$ and $c_x=c_z>0$. The system harbors both  shifted first-order nodal lines and third-order Fermi-arcs consisting of zero-mode corner states of 3D TOTIs { located at $(x,y,z)=(\mp L/2,\pm L/2,\mp L/2)$} stacked along $k_w$ axis. }
	\label{FC3}
\end{figure}
 
 We first consider the situation when $\Delta_2$ plays the role in, i.e., all $c_i=0$. $\Delta_2$ 
 breaks $\T$, $\C$, but keeps $\S$, $\I$ and thus the system is now falls into calss AIII but still hosts an unchanged spin SCN. The discussion of the second-order topology associated with the spin MFCNs \cite{N8} is similar to the cases in $\H_1$ when only $b_{xw}$ and $b_{yw}$ survive\cite{SM}. To conveniently discuss third-order topology later, we focus on a TI phase with both first- and second-order boundary states with $b_{xw}=b_{yw}=b_{zw}$ for $M$ near $3$.  As shown in Fig. \ref{FC3}(a-b), the system hosts the second-order helical Fermi-arcs  where its 3D layers $\H_2^{k_w}$ with $k_w\in (-b_{xw},b_{xw})$ are SOTIs with helical states propagating along $\M_{x-y}$, $\M_{y-z}$, and $\M_{z-x}$ symmetric hinges\cite{N1} when considering OBCs along $x$-, $y$- and $z$-directions. One can also define 3D mirror winding numbers to describe this second-order topology accordingly\cite{SM}.

 Subsequently, we turn on $c_{x}=c_z$ being numeric small, $\H_3$ remains in class AIII harboring shifted 3D $\Z_2$ nodal lines \cite{LBShao2018,JXDai2020,Ahn2018} on the 3D boundaries when assuming OBC along $x$ or $y$ directions [Fig. \ref{FC3}(c)]. However, helical hinge Fermi-arcs  have been gapped due to $\Delta_3$ breaks $\tilde{\Gamma}_5$ ($\M_{x-y}$ and $\M_{y-z}$ are also broken), leading to third-order Fermi-arcs along $k_w$ axis\cite{SM}, as shown in Fig. \ref{FC3}(c-e).  In this case, even though the spin SCN is not well-defined, one can define a $\Z_2$ number as 4D generalization of Kane-Mele invariant \cite{LBShao2018} since $\H_3$ remains respect $\R_w\T$-symmetry\cite{N6}.  Alternately, we can define another topological number as $\nu=\nu_{0}-\nu_{\pi}~\text{mod}~2$ at $\R_w\T$-invariant lines $k_w=0,\pi$ with $\nu_{k_w}$ is a topological number of the 3D real TI with $\S$ \cite{JXDai2021}. Calculation results show that $\nu_0=1$ and $\nu_{\pi}=0$ for $M$ near $3$ \cite{SM}.
Meanwhile, $\H_3$ keeps $\M_{(z+x)yw}=\M_{z+x}\M_y\M_w$\cite{N7}, $\M_{z-x}$, and $\I$-symmetries.  A non-trivial 1D mirror winding number $w_{1m}=2$ defined at $k_x=-k_z$ and $(k_y,k_w)=(0,0)$ associated with $\M_{(z+x)yw}$-symmetry \cite{SM} guarantees the exact location of third-order Fermi-arc with $\M_w$-symmetry about $k_w=0$ consisting of a pair of zero-modes, accumulating at two $\M_{(z+x)y}$-symmetric [combined mirror reflection about $z=-x$ and $y=0$ planes] corners in a 3D real space cubic spanned in $(x,y,z)$. The bulk-corner correspondence is established here; $\M_{z-x}$ implies these two zero-modes live in $z=x$ plane; these Fermi-arc states are always $\I$-symmetric in $(x,y,z,k_w)$ space [as seen in Fig. \ref{FC3}(d-e)].  We emphasize that $\tilde{\Gamma}_5$ $(\R_w\T)$ symmetry guarantees the existence of second (third)-order gapless Fermi arcs in class AIII, where these boundary modes are located at the positions determined by additional crystalline symmetries.

\begin{figure}[http]\centering
	\includegraphics[width=8cm]{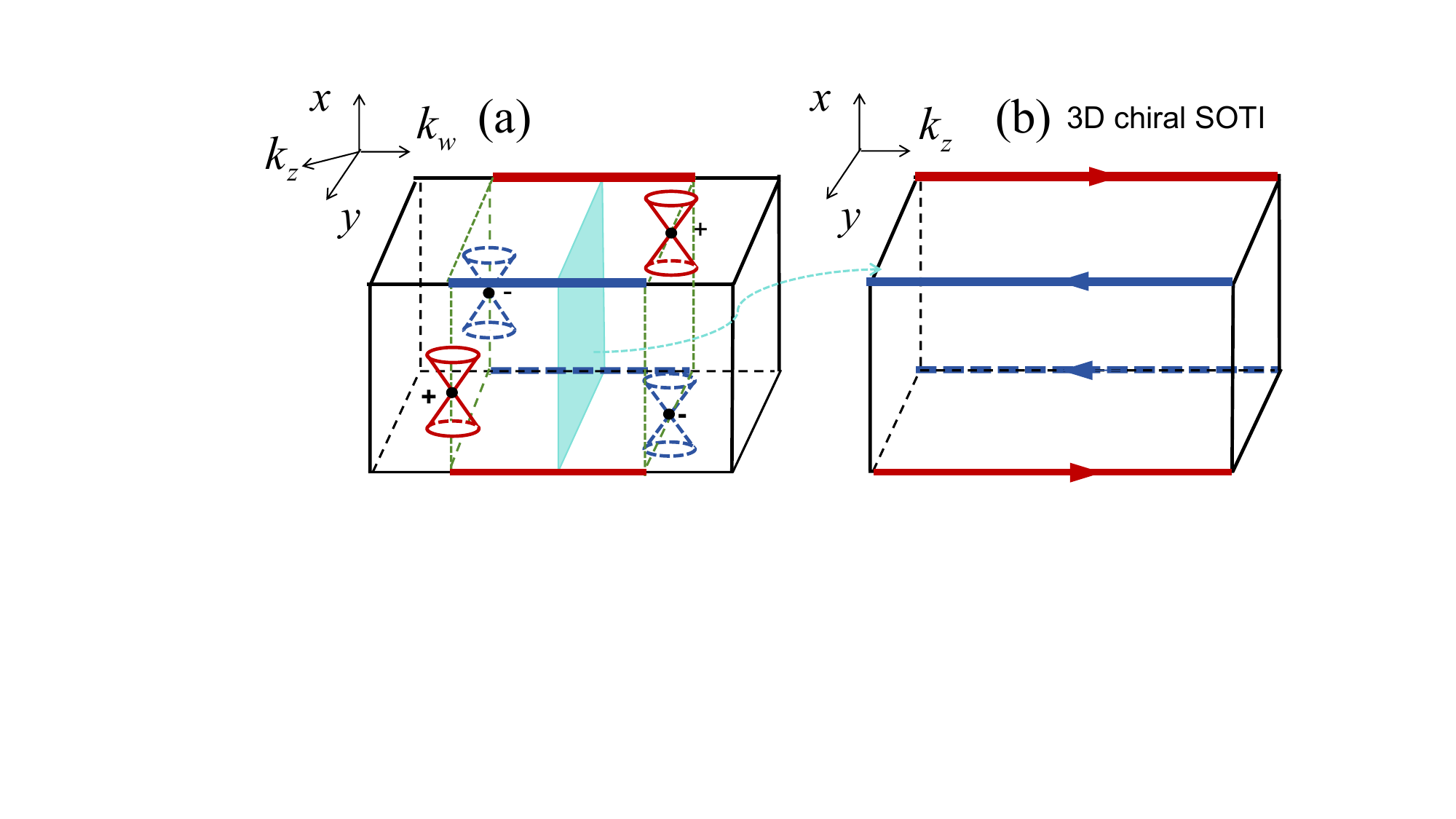}
	\caption{ Schematic of the boundary states of $\H_3$  when $2<M<4$ and $\delta>0$ being numeric small.  (a) The system harbors both  shifted first-order Weyl cones \cite{SM} and the second-order chiral Fermi-arc hinge states. The green dashed squares denote the 3D TCIs being the  unconventional phase transition points. {(b) 3D non-trivial SOTI hosts second-order chiral states located at four hinges.}    }
	\label{FC2}
\end{figure}

\emph{\color{blue}Four-band model in Class A$^{C_4\T}$ and eight-band model in Class AIII$^{C_4\T}$ without SOC.---}
We now address another 4D four-band model in class A, with the Hamiltonian,
\begin{equation}
	\H_3(\boldsymbol k)=\H_0+\Delta_4,
\end{equation}
with a perturbation $\Delta_{4}=\delta(\cos k_x-\cos k_y)\Gamma_4$ breaks  $C_4^{zw}$ and $\T$
individually, but respects $C_4^{zw}\T$-symmetry \cite{CT}.
As shown in Fig. \ref{FC2}, $\Delta_4$ shifts the boundary Weyl cones along $k_w$-direction \cite{SM}. Two  Weyl cones located on different boundaries [e.g. $y(x)=-L/2$ and $y(x)=L/2$] are shifted along the same direction when OBC along $y$ ($x$)-direction is assumed, resulting in a second-order Fermi-arc at four hinges once we further open the boundary along $x$ $(y)$ axis.  Except the unchanged $C_2$ for the first-order topology, we can define the so-called ``\emph{rotational first Chern number}" (RFCN) at $C_4^{zw}\T$-invariant 2D momenta $(k_x,k_y)=(0,0),(\pi,\pi)$ which may be used to characterize the second-order topology \cite{SM}.  
In addition,
we can easily construct the spin conservation verion of $\H_3$ with eight-band as $\H_4(\boldsymbol k)=\sigma_{3}\otimes\H_3$, which respects $C^{zw}_4\T$ and $\S$ and thus falls into class AIII.  The system now hosts the first-order boundary Dirac modes and the second-order helical hinge Fermi-arcs. By setting $k_z=0$, we obtain the corresponding 3D unique TI. All these similar discussions about the second-order topology associated with additional crystalline symmetries are presented in Ref.\cite{SM}.

\emph{\color{blue}CTPTs and Novel TSMs.---}By enlarging the parameters in $\Delta_1/\Delta_2$ of $\H_1$/$\H_2$, we can study the CTPT goes through a phase transition from  this unique TI phase to a non-trivial nodal-line/sphere metallic phase and then finally becomes a trivial insulator. Such 4D nodal structures, including  nodal-line, spin-nodal-line, and  nodal-sphere, are characterized by the first Chern number (FCN) $C_1$, spin FCN $C_{1s}$, and 1D winding number $w_1$ respectively. The corresponding topological responses based on current works \cite{CWang2021,Ramamurthy2017,Hirsbrunner2023} are also discussed \cite{SM}.
On the other hand, we can obtain serveral new classes of TSMs with hybrid-order boundary Fermi-arcs/surfaces by stacking the above TI phases along an extra spatial dimension $u$, as is commonly done, i.e., $M\rightarrow M'=M-\cos k_u$,   For instance, when $M=4$, $\H_1(M')$/$\H_1^{k_z=0}(M')$ hosts a pair of 5D/4D nodal-surfaces/lines characterized by double charges $\nu_D=(C_2/w_3,\Delta C_1)$, where $w_3$ denotes the 3D winding number; $\H_3(M')$/$\H_3^{k_z=0}(M')$ hosts a pair of hybrid-order 5D/4D monopoles characterized by $C_2/w_3$. Two spin copies of these nodal-structures of $\H_2(M')/\H_4(M')$ can be also extended accordingly\cite{SM}.

\emph{\color{blue}Discussion and outlook.---}
In this work, we have established a valuable framework {in which these unique TIs can be viewed as a higher-dimensional stacking effect of low-dimensional HOTIs undergoing UTPTs along an additional spatial dimension, helping to elucidate their topological origin. These novel TIs, induced from first-order TIs, exhibit distinctive properties, including first-order topological boundary modes and higher-order Fermi arcs, which are presented for the first time in this work. In  general},  each kind of $d$D HOTI corresponds to a $(d+1)$D unique TI phase for $d\geq 2$.  The related nodal structures and their topological responses are also studied when $\H_1/\H_2$ becomes a gapless phase. Furthermore, the corresponding TSMs with novel nodal-structures can be obtained by regularly stacking these TIs along an extra spatial dimension.
Note that these unique topological phases,  especially for models with $d\geq 4$,  could
be experimental studied in sythetic matter, such as metamaterials \cite{WCheng2021,SMa2021,ZGChen2021,HChen2021},  electric circuits \cite{YWang2020,RYu2020,ZWang2023}, superconducting circuits \cite{XSTan2018,XSTan2021,YZhang2024}, diamond with nitrogen vacancy center \cite{FKong2016,MYu2019,MChen2022}, or cold atomic systems \cite{SLZhu2006,SLZhu2007,LShao2008,Price2015,Goldman2014a,YQZhu2017,Petrides2018,Lohse2018,Bouhiron2024}, etc. In particular, we can utilize the cold-atom setup proposed in our previous work \cite{YQZhu2022} to realize the 4D model $\tilde{\H}_0$ in Eq. \eqref{4Dmodel2}, including the implementation of other models mentioned in this work in a similar manner \cite{SM}.
For the 3D model in $\H_1$ with $k_z=0$, $\H_0^{k_z=0}$ as a 3D strong TI in low-energy limit has been predicted \cite{HZhang2009} and  realized \cite{YLChen2009} in real materials, such as Bi$_2$Se$_3$ and Bi$_2$Te$_3$ families, one can consider to induce $\Delta_1$ term in these materials by magnetization or light-induced Floquet drivings \cite{N10}.

We highlight that these novel topological phases can also be generalized to non-Hermitian \cite{HShen2018,ZGong2018,SYao2018} and Floquet systems \cite{Cayssol2013,Goldman2014,Roy2019a,Salerno2020}. The field-theoretical descriptions of these phases in the bulk is one of interesting directions in the future. Overall, our work further opens the door of investigating novel topological phases in the synthetic and real materials.

\begin{acknowledgements}
This work was supported by the National Key R\&D Program of China (Grants No. 2024YFA1409300, No. 2022YFA1405304), the Guangdong-Hong Kong Joint Laboratory of Quantum Matter, the NSFC/RGC JRS grant (Grant No. N\underline{~~}HKU774/21),  the CRF (Grant No. C6009-20G) and  GRF (Grants No. 17310622 and No. 17303023) of Hong Kong, Guangdong Provincial Quantum Science Strategic Initiative (GDZX2200001),  and Open Fund of Key Laboratory of Atomic and Subatomic Structure and Quantum Control (Ministry of Education).
\end{acknowledgements}

\bibliographystyle{apsrev4-1}
\bibliography{myref}

\begin{thebibliography}{104}
\bibitem [{SCN()}]{SCN}%
  \BibitemOpen
\bibfield  {journal} {  }\href@noop {} {\bibinfo  {journal} {Direct calculation
  for $\H_0$ shows that $C_2=-\text{sign}(M)$ for $2<|M|<4$ and
  $C_{2}=3\text{sign}(M)$ for $0<|M|<2$, and $C_2=0$ elsewhere}\ }\BibitemShut
  {NoStop}%
\bibitem [{MS()}]{MS}%
  \BibitemOpen
\bibfield  {journal} {  }\href@noop {} {\bibinfo  {journal}
  {$\M_i:k_i\rightarrow -k_i$ for $i\in (x,y,z,w)$. The operator of
  $\mathcal{M}_{ij}$ is represented as
  $\hat{\mathcal{M}}_{ij}=\Gamma_{i}\Gamma_j$ with
  $(\hat{\mathcal{M}}_{ij})^2=-1$}\ }\BibitemShut {NoStop}%
\bibitem [{HSP()}]{HSP}%
  \BibitemOpen
\bibfield  {journal} {  }\href@noop {} {\bibinfo  {journal} {The high-symmetry
  points
  $\Lambda_a^{xw/yz}=(k_x,k_w)/(k_y,k_z)=(0,0),(0,\pi),(\pi,0),(\pi,\pi)$ are
  labeled as $a=1,2,3,4$ in sequential order}\ }\BibitemShut {NoStop}%
\bibitem [{N4()}]{N4}%
  \BibitemOpen
\bibfield  {journal} {  }\href@noop {} {\bibinfo  {journal} {The symmetry
  operators are $\TT=G_{002}\K$, $\CC=G_{102}\K$, and $\SS=\tilde{\Gamma}_{6}$
  satisfying $\TT^2=-1$, $\CC^2=-1$, and $\SS^2=+1$}\ }\BibitemShut {NoStop}%
\bibitem [{N5()}]{N5}%
  \BibitemOpen
\bibfield  {journal} {  }\href@noop {} {\bibinfo  {journal} {When the system
  hosts $\tilde{\Gamma}_5$ symmetry, we can define a $\Z_2$ number through the
  spin SCN as $\nu_2=C_{2s}~\text{mod}~2$ with $C_{2s}=(C_2^{+}-C_2^-)/2$ where
  $C_2^{\pm}$ is the SCN for each $4\times 4$ block
  Hamiltonian\cite{YQZhu2022}}\ }\BibitemShut {NoStop}%
\bibitem [{N8()}]{N8}%
  \BibitemOpen
\bibfield  {journal} {  }\href@noop {} {\bibinfo  {journal} {The spin MFCNs are
  defined as $C_{1sm}^{ij}=(C_{1m,+}^{ij}-C_{1m,-}^{ij})/2$ for $ij=xw,yz$ with
  $C_{1m,\pm}^{ij}$ defined in Eq. \eqref{MC} of each $4\times 4$ block
  Hamiltonian, respectively}\ }\BibitemShut {NoStop}%
\bibitem [{N7()}]{N7}%
  \BibitemOpen
  \href@noop {} {\bibinfo  {journal} {The representative operator of
  $\M_{(x+z)yw}$ is given by
  $\MM_{(z+x)yw}=\MM_{z+x}\MM_y\MM_w=\frac{1}{\sqrt{2}}(G_{223}-G_{221})$ with
  $(\MM_{(z+x)yw})^2=+1$}\ }\BibitemShut {NoStop}%
\end{thebibliography}

\widetext
\clearpage


\section*{Supplementary material}

# Supplemental Materials for "Novel Topological Insulators with Hybrid-order Boundary States"

Yan-Qing Zhu,[1,2,*] Zhen Zheng,[3,4] Giandomenico Palumbo,[5,†] and Z. D. Wang[1,2,‡]

[1]*Guangdong-Hong Kong Joint Laboratory of Quantum Matter, Department of Physics,*
*and HK Institute of Quantum Science & Technology,*
*The University of Hong Kong, Pokfulam Road, Hong Kong, China*
[2]*Quantum Science Center of Guangdong-Hong Kong-Macau Great Bay Area, 3 Binlang Road, Shenzhen, China*
[3]*Key Laboratory of Atomic and Subatomic Structure and Quantum Control (Ministry of Education),*
*Guangdong Basic Research Center of Excellence for Structure and Fundamental Interactions of Matter,*
*School of Physics, South China Normal University, Guangzhou 510006, China*
[4]*Guangdong Provincial Key Laboratory of Quantum Engineering and Quantum Materials,*
*Guangdong-Hong Kong Joint Laboratory of Quantum Matter,*
*Frontier Research Institute for Physics, South China Normal University, Guangzhou 510006, China*
[5]*School of Theoretical Physics, Dublin Institute for Advanced Studies, 10 Burlington Road, Dublin 4, Ireland*
(Dated: May 27, 2024)

In this file, we first provide definitions of the relevant symmetries mentioned in the paper. We then proceed to a detailed discussion of the models mentioned in the main text, including the boundary Hamiltonian, calculating the topological numbers associated with the corresponding crystalline symmetries, and analyzing the topology of the nodal-line/sphere phases of $\mathcal{H}_1$ and $\mathcal{H}_2$, as well as providing their field-theoretical descriptions. Subsequently, we explore the construction of corresponding novel topological semimetals in higher-dimensions by stacking these novel TIs along an additional spatial dimension. Finally, we provide a feasible proposal for implementing the four-band models discussed in the main text, whereas the proposal for the eight-band model can be found in our previous work.

* zhuyq1992@gmail.com
† giandomenico.palumbo@gmail.com
‡ zwang@hku.hk

## CONTENTS

## S-I. $\mathcal{T}$, $\mathcal{C}$, $\mathcal{S}$ AND CRYSTALLINE SYMMETRIES

The well-known time-reversal, charge-conjugate, and chiral (sublattice) symmetries, acting on the Bloch Hamiltonian as,

$$
\begin{aligned}
\mathcal{T}\mathcal{H}(\boldsymbol{k})\mathcal{T}^{-1} &= \mathcal{H}(-\boldsymbol{k}), \\
\mathcal{C}\mathcal{H}(\boldsymbol{k})\mathcal{C}^{-1} &= -\mathcal{H}(-\boldsymbol{k}), \\
\mathcal{S}\mathcal{H}(\boldsymbol{k})\mathcal{S}^{-1} &= \mathcal{H}(\boldsymbol{k}),
\end{aligned}
\tag{S1}
$$

respectively, with the aniunitary $\mathcal{T}^2 = \pm 1$, $\mathcal{C}^2 = \pm 1$, and the unitary $\mathcal{S}^2 = +1$ forming the ten-fold AZ classification [1].

We next discuss the crystalline symmetries presented in our work about the 4D models. For a system $\mathcal{H}(\boldsymbol{k})$ preserves a crystalline symmetry $\mathcal{G}$, we have

$$\mathcal{G}\mathcal{H}(\boldsymbol{k})\mathcal{G}^{-1} = \mathcal{H}(g\boldsymbol{k}),$$ (S2)

where we take $\mathcal{G}$ to be unitary and $g$ tranforms the momentum $\boldsymbol{k}$. The first one is the inversion symmetry $\mathcal{I}$, acting on the Hamiltonian as, $\mathcal{I}\mathcal{H}(\boldsymbol{k})\mathcal{I}^{-1} = \mathcal{H}(-\boldsymbol{k})$. In other words, $\mathcal{I}$ inverses all the momentum components and its dual coordinate in real space:

$$\begin{aligned} \mathcal{I} : \boldsymbol{k} = (k_x, k_y, k_z, k_w) \to -\boldsymbol{k} = (-k_x, -k_y, -k_z, -k_w), \\ \mathcal{I} : \boldsymbol{r} = (x, y, z, w) \to -\boldsymbol{r} = (-x, -y, -z, -w), \end{aligned}$$ (S3)

with $\mathcal{I}$ being the unitary transformation. If the unitary tranformation only reflection one of the spatial dimension components, then we have a so-called mirror reflection symmetry $\mathcal{M}_i$ with a mirror plane at $k_i/i$ direction, i.e.,

$$\mathcal{M}_i \mathcal{H}(k_i, k_j, k_l, k_m) \mathcal{M}_i^{-1} = \mathcal{H}(-k_i, k_j, k_l, k_m),$$ (S4)

for $(i, j, l, m) \in (x, y, z, w)$ hereafter. In other words, $\mathcal{M}_i : k_i \to -k_i \leftrightarrow i \to -i$. For instance, when we take $i = x$, $\mathcal{M}_x$-invariant planes $k_x = 0$ and $k_x = \pi$ correspond to the mirror planes $x = 0$ and $x = a/2$ with $a$ being the lattice spacing in real space respectively [2]. For convenience, we take all the representation of $\mathcal{M}_i$ as $(\hat{\mathcal{M}}_i)^2 = -1$. Based on this point, we can further define a combined mirror reflection for two of the spatial components, i.e., $\mathcal{M}_{ij} = \mathcal{M}_i\mathcal{M}_j$, acting as,

$$\mathcal{M}_{ij}\mathcal{H}(\boldsymbol{k})\mathcal{M}_{ij}^{-1} = \mathcal{H}(g\boldsymbol{k}),$$ (S5)

where $g\boldsymbol{k} = (-k_i, -k_j, k_l, k_m)$ with $\mathcal{M}_{ij}^2 = -1$. For this case, we have four $\mathcal{M}_{ij}$-invariant points at $(k_i, k_j) = (0, 0)$, $(0, \pi)$, $(\pi, 0)$, and $(\pi, \pi)$. Moreover, one can further define a combined mirror reflection of three of the spatial components as $\mathcal{M}_{ijk}$ with $g\boldsymbol{k} = (-k_i, -k_j, -k_l, k_m)$ with eight $\mathcal{M}_{ijk}$-invariant points.

In what follows we discuss the four-fold rotation in the $i$-$j$ plane about $lm$-axis with counterclockwise as the positive direction, i.e.,

$$C_4^{lm}\mathcal{H}(\boldsymbol{k})(C_4^{lm})^{-1} = \mathcal{H}(g\boldsymbol{k}),$$ (S6)

with $g\boldsymbol{k} = (-k_j, k_i, k_l, k_m)$. With the definition of $C_4^{lm}$ and $\mathcal{M}_i$, we can define a mirror reflection $\mathcal{M}_{i\pm j}$ about reflection-invariant plane about $k_i = \mp k_j$, acting as,

$$\mathcal{M}_{i\pm j}\mathcal{H}(k_i, k_j, k_l, k_m)\mathcal{M}_{i\pm j}^{-1} = \mathcal{H}(\mp k_j, \mp k_i, k_l, k_m).$$ (S7)

Similarly, in real space, $\mathcal{M}_{i\pm j}$: $(i, j, l, m) \to (\mp j, \mp i, l, m)$. These reflections can be decoupled as two combined operations as $\mathcal{M}_{i+j} = \mathcal{M}_j C_4^{lm}$ and $\mathcal{M}_{i-j} = \mathcal{M}_i C_4^{lm}$. For instance, $\mathcal{M}_{x-y} = \mathcal{M}_x C_4^{zw}$, it reflects $(x, y)$ to $(y, x)$ about the mirror plane at $x = y$ in real space. One can define the related topological number at $k_x = k_y$ plane for the eigenspace of $\mathcal{M}_{x-y}$ in $\boldsymbol{k}$ space to characterize the higher-order topology; see the following sections in concrete models. In general, one can also further define a combined symmetry with $\mathcal{M}_{i\pm j}$ and $\mathcal{M}_l$, etc.

When considering the combination of $\mathcal{T}$ and $\mathcal{G}$, we have an antiunitary combined symmetry. For instance, $C_4^{lm}\mathcal{T}$ acts on $\mathcal{H}$ as,

$$C_4^{lm}\mathcal{T}\mathcal{H}(k_i, k_j, k_l, k_m)(C_4^{lm}\mathcal{T})^{-1} = \mathcal{H}(k_j, -k_i, -k_l, -k_m),$$ (S8)

where $(C_4^{lm}\mathcal{T})^4 = -1$. Another combined symmetry mentioned in this work is $\mathcal{R}_i\mathcal{T}$-symmetry, acting on the Hamiltonian as,

$$\mathcal{R}_i\mathcal{T}\mathcal{H}(k_i, k_j, k_l, k_m)(\mathcal{R}_i\mathcal{T})^{-1} = \mathcal{H}(-k_i, k_j, k_l, k_m),$$ (S9)

where $\mathcal{R}_i\mathcal{T} = \mathcal{M}_i\mathcal{I}\mathcal{T}$ with $(\mathcal{R}_i\mathcal{T})^2 = +1$.

## S-II. COMMON TOPOLOGICAL NUMBERS IN 1D/2D/3D/4D

In this section, we present the definitions of common topological numbers in different spatial dimensions which are mentioned in the main text or in the following sections associated with the new topological numbers under certain crystalline symmetries.

### A. Zak phase and 1D winding number

For a 1D topological insulator $\mathcal{H}(k)$ protected by certain symmteries, such as $\mathcal{S}$, $\mathcal{C}$, etc, we can calculate the quantized Zak phase of this system with the form,

$$\gamma_{Zak} = \int_{-\pi}^{\pi} dk \ \text{tr} \ \mathcal{A}_k, \tag{S10}$$

where the non-Abelian Berry connection $\mathcal{A}_k = \langle \psi | i \partial_k | \psi \rangle$ with its component $\mathcal{A}_k^{mn} = \langle \psi_m | i \partial_k | \psi_n \rangle$. Here $|\psi_n\rangle$ denotes the eigenstate of the occupied bands.

Especially, if $\mathcal{H}(k)$ respects $\mathcal{S}$, one can define a 1D winding number $w_1$ as $w_1 = \gamma_{Zak}/\pi$ or

$$w_1 = \frac{1}{4\pi i} \int_{-\pi}^{\pi} dk \text{tr} \left( \hat{\mathcal{S}} \mathcal{H} \partial_k \mathcal{H}^{-1} \right) \tag{S11}$$

where $\hat{\mathcal{S}}$ is the representative operator of chiral symmetry[3].

### B. First Chern number

For a 2D topological insulator $\mathcal{H}(k_i, k_j)$, we can calculate the first Chern number as

$$C_1 = \frac{1}{2\pi} \int_{T^2} d^2 k \ \text{tr} \ \mathcal{F}_{ij} \tag{S12}$$

where $\mathcal{F}_{ij} = \partial_{k_i} \mathcal{A}_j - \partial_{k_j} \mathcal{A}_i - i[\mathcal{A}_i, \mathcal{A}_j]$ with $\mathcal{A}_i$ is defined for the occupied bands of $\mathcal{H}$ as above. Here $T^2$ denotes 2D Brillouin Zone (BZ).

### C. Chern-Simons form and 3D winding number

For a 3D topological insulator $\mathcal{H}(k_i, k_j, k_l)$, one can calculate the theta angle or Chern-Simons form [1] as,

$$\Theta = \frac{1}{4\pi} \int_{T^3} d^3 k \epsilon^{ijl} \text{tr} \left( \mathcal{A}_i \partial_{k_j} \mathcal{A}_l - \frac{2i}{3} \mathcal{A}_i \mathcal{A}_j \mathcal{A}_l \right), \tag{S13}$$

which is qunatized by the certain symmetries, such as $\mathcal{T}$, $C_4\mathcal{T}$, etc, taking the vaule $\Theta = 0, \pi$. Here $\epsilon^{ijl}$ denotes the Levi-Civita symbol and $T^3$ the 3D BZ.

If the system preserves symmetry $\mathcal{S}$, we can define a 3D winding number as $\Theta/\pi$, or,

$$w_3 = -\frac{1}{48\pi^2} \int_{T^3} d^3 k \epsilon^{ijl} \text{tr} \left( \hat{\mathcal{S}} \mathcal{H} \partial_{k_i} \mathcal{H}^{-1} \mathcal{H} \partial_{k_j} \mathcal{H}^{-1} \mathcal{H} \partial_{k_i} \mathcal{H}^{-1} \right). \tag{S14}$$

For a Dirac-type Hamiltonian $\mathcal{H} = d_1 \Gamma_1 + d_2 \Gamma_2 + d_3 \Gamma_3 + d_0 \Gamma_0$ with Dirac matrices $\{\Gamma_i, \Gamma_j\} = 0$, the expression can be further simplied as

$$w_3 = \frac{1}{12\pi^2} \int_{T^3} d^3 k \frac{1}{d^4} \epsilon^{\mu\nu\alpha\beta} \epsilon^{ijk} d_\mu \partial_i d_\nu \partial_j d_\alpha \partial_k d_\beta, \tag{S15}$$

with $d = \sqrt{d_1^2 + d_2^2 + d_3^2 + d_0^2}$, $\mu \in \{1, 2, 3, 4\}$, $i \in \{x, y, z\}$, and $\partial_i \equiv \partial_{k_i}$.

### D. Second Chern number

For a 4D topological insulator, one can define the second Chern number [4] as

$$C_2 = \frac{1}{32\pi^2} \int_{T^4} d^4 k \ \epsilon^{\mu\nu\rho\sigma} \text{tr}(\mathcal{F}_{\mu\nu} \mathcal{F}_{\rho\sigma}), \tag{S16}$$

where $(\mu, \nu, \rho, \sigma) \in (x, y, z, w)$. Direct calculation for $\mathcal{H}_0$ shows that $C_2 = -\text{sign}(M)$ for $2 < |M| < 4$ and $C_2 = 3\text{sign}(M)$ for $0 < |M| < 2$, and $C_2 = 0$ elsewhere. Here $T^4$ denotes 4D BZ.

## S-III. $\mathcal{H}_1$ IN CLASS A$^{\mathcal{M}_{ij}}$ AND $\mathcal{H}_1^{k_z=0}$ IN CLASS AIII$^{\mathcal{M}_i}$

### A. $\mathcal{H}_1$ in Class A$^{\mathcal{M}_{ij}}$

#### 1. Shifted boundary Weyl cones

Consider the 4D model $\mathcal{H}_1 = \mathcal{H}_0 + \Delta_1$ in the main text, where

$$\mathcal{H}_0 = \sin k_x \Gamma_1 + \sin k_y \Gamma_2 + \sin k_z \Gamma_3 + \sin k_w \Gamma_4 + (M - \sum_{i=x,y,z,w} \cos k_i)\Gamma_0, \tag{S17}$$

and

$$\Delta_1 = b_{xw} \Gamma_0 \Gamma_{14} + b_{yw} \Gamma_0 \Gamma_{24}. \tag{S18}$$

The projectors upon applying open boundary condition along $\alpha$-direction are given by [5, 6]

$$P_\alpha^\pm = \frac{1}{2}(1 \mp i\Gamma_\alpha \Gamma_0), \tag{S19}$$

where $\pm$ represents the projection of $\mathcal{H}_1$ to the boundary at $\alpha = \pm L/2$, with a lattice length of $L$. Here we mix the notation in the subscript of $\Gamma_\alpha$: $\alpha = (x, y, z, w) \leftrightarrow (1, 2, 3, 4)$.

Thus the effective Hamiltonian at the first-order boundaries for $\mathcal{H}_1$ is given by,

$$\mathcal{H}_{\alpha,eff}^\pm = P_\alpha^\pm \mathcal{H}_1 P_\alpha^\pm, \tag{S20}$$

for $|\lambda_\alpha| < 1$ where

$$\lambda_\alpha = M - \sum_{i,i\neq\alpha} \cos k_i. \tag{S21}$$

In the diagonal representation of the projectors, i.e., $u_\alpha P_\alpha^+ u_\alpha^{-1} = 1_2 \oplus 0_2$, and $u_\alpha P_\alpha^- u_\alpha^{-1} = 0_2 \oplus 1_2$ with $1_2$ denotes the $2 \times 2$ identity matrix and $0_2$ is zero matrix in a $2 \times 2$ representation, we have all boundary effective Hamiltonian, given by,

$$\begin{aligned}
\mathcal{H}_{x,eff}^\pm &= \mp \sin k_y \sigma_2 \mp \sin k_z \sigma_3 \pm (\sin k_w \pm b_{xw})\sigma_1, \\
\mathcal{H}_{y,eff}^\pm &= \mp \sin k_x \sigma_1 \mp \sin k_z \sigma_3 \pm (\sin k_w \pm b_{yw})\sigma_2, \\
\mathcal{H}_{z,eff}^\pm &= \mp \sin k_x \sigma_1 \mp \sin k_y \sigma_2 \pm \sin k_w \sigma_3, \\
\mathcal{H}_{w,eff}^\pm &= \mp(\sin k_x \mp b_{xw})\sigma_1 \mp (\sin k_y \mp b_{yw})\sigma_2 \mp \sin k_z \sigma_3,
\end{aligned} \tag{S22}$$

with $u_x = e^{i\frac{\pi}{4}G_{11}}$, $u_y = e^{i\frac{\pi}{4}G_{12}}$, $u_z = e^{i\frac{\pi}{4}G_{13}}$, and $u_w = 1_4$ being the identity.

As we can see that there is a shifted Weyl cone at $x/y = \pm L/2$ along $k_w$ axis for $2 < M < 4$, whose low-energy Hamiltonian near the origin is given by

$$\begin{aligned}
\mathcal{H}_{x,\pm} &= \mp k_y \sigma_2 \mp k_z \sigma_3 \pm (k_w + b_{xw})\sigma_1, \\
\mathcal{H}_{y,\pm} &= \mp k_x \sigma_1 \mp k_z \sigma_3 \pm (k_w + b_{yw})\sigma_2,
\end{aligned} \tag{S23}$$

respectively. At the boundary $x = \pm L/2$, there is a Weyl cone located at $\mathbf{k}_{x\pm} = (k_y, k_z, k_w) = (0, 0, \mp b_{xw})$ with chirality $\chi_\pm = \pm 1$; At $y = \pm L/2$, there is a Weyl cone located at $\mathbf{k}_{y\pm} = (k_x, k_z, k_w) = (0, 0, \mp b_{yw})$ with chirality $\chi_\pm = \pm 1$. Note that $b_{xw}/b_{yw}$ both acts on the boundaries normal to $x/y$ and $w$ direction, shifting the Weyl cones along $k_w$ and $k_x/k_y$ directions, respectively.

#### 2. Calculation of Mirror first Chern number

When only $b_{xw}$ is non-zero in $\Delta_1$, $\mathcal{H}_1$ preserves $\mathcal{M}_{xw}$ and $\mathcal{M}_{yz}$ symmetries acting as in Eq. (S5). For $\mathcal{M}_{xw}$-symmetry, we define mirror first Chern number (MFCN) $C_{1m}^{xw}$ at four high symmetric points $\Lambda_1^{xw} = (0, 0)$, $\Lambda_2^{xw} = (0, \pi)$,

$\Lambda_3^{xw} = (\pi, 0)$, $\Lambda_4^{xw} = (\pi, \pi)$. By digonalizing $\hat{\mathcal{M}}_{xw} = \Gamma_1\Gamma_4$ into $iG_{30}$ with $U_{xw} = e^{-i\frac{\pi}{4}G_{21}}$, we have the corresponding Hamiltonian at $\Lambda_a^{xw}$, given by,

$$U_{xw}\mathcal{H}_1(\Lambda_a^{xw})U_{xw}^{-1} = \sin k_y G_{32} + \sin k_z G_{33} - (m_a - \cos k_y - \cos k_z)G_{31} + b_{xw}G_{01} = \mathcal{H}_+ \oplus \mathcal{H}_-, \tag{S24}$$

where $m_a = M - \cos k_x - \cos k_w$ at $\Lambda_a^{xw}$ being

$$m_1 = M - 2, m_2 = M, m_3 = M, m_4 = M + 2, \tag{S25}$$

respectively. We below give the phase diagram of this 2D model in Eq. (S50), which phase transition points are now given by $m_a \pm b_{xw} = 0, \pm 2$, then we have six phase boundaries,

$$b_{xw} = \pm m_a, -m_a \pm 2, m_a \pm 2. \tag{S26}$$

When $b_{xw} = 0$, we have the phase diagram for $\mathcal{H}_\pm$ labeling with the first Chern number $C_{\pm i}$ as a function of $m_a$. After tuning on $b_{xw}$, effective mass $\tilde{m}_{a,\pm} = m_a \pm b_{xw}$, one can easily obtain $C_{\pm i}$ fo $\mathcal{H}_\pm$ by shifiting the parameter region and obtain the full phase diagram in Fig. S1. When $b_{xw}$ being numeric small, $\mathcal{H}_\pm$ with mass term $\tilde{m}_{a,\pm} = m_a \pm b_{xw}$ hosts opposite first Chern number as defined in Eq. (S12), i.e., $C_{+i} + C_{-i} = 0$, thus we can define MFCN as

$$C_{1m}^{xw} = (C_{+i} - C_{-i})/2, \tag{S27}$$

for the gapped phase e.g., $m_a \in (0, 2) \cup (-2, 0)$ and $-1 < b_{xw} < 1$.

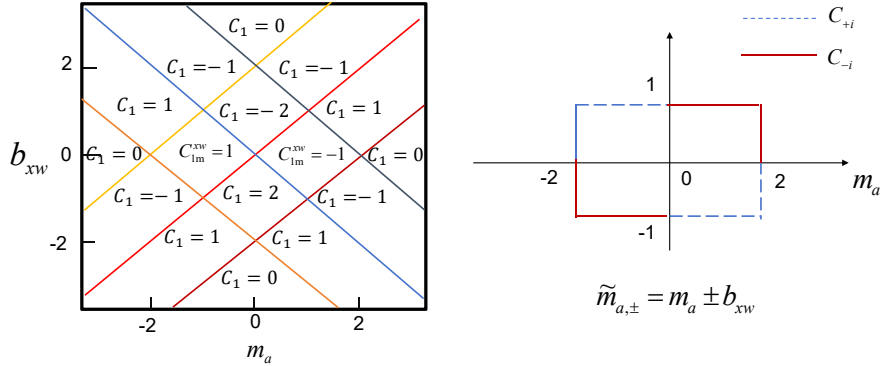

FIG. S1. Left panel: Schematic of the phase diagram of $\mathcal{H}_1(\Lambda_a^{xw})$ with parameters $(m_a, b_{xw})$. The straight color lines denote the phase transition boundaries where the bulk gap closes at $E = 0$. Note that $C_1 = C_{+i} + C_{-i}$. Right panel: The schematic of the phase diagram of $\mathcal{H}_\pm$ as the conventional 2D Chern insulators with $C_{\pm i}$ by varying $m_a$ when $b_{xw} = 0$.

Now back to our 4D gapped case in the region for $2 + |b_{xw}| < M < 4 - |b_{xw}|$ with $0 < |b_{xw}| < 1$, we have

$$C_{1m}^{xw}(\Lambda_1^{xw}) = -1, C_{1m}^{xw}(\Lambda_i^{xw}) = 0, i = 2, 3, 4. \tag{S28}$$

For $|b_{xw}| < M < 2 - |b_{xw}|$ with $0 < |b_{xw}| < 1$, we have

$$C_{1m}^{xw}(\Lambda_1^{xw}) = 1, C_{1m}^{xw}(\Lambda_{2/3}^{xw}) = -1, C_{1m}^{xw}(\Lambda_4^{xw}) = 0. \tag{S29}$$

In this case there are three pairs of second-order hinge Fermi-arcs with the location with mirror-reflection about $x = 0$ and $k_w = 0$, $x = a/2$ and $k_w = 0$, $x = 0$ and $k_w = \pi$, respectively. The discussion associated with $C_{1m}(\Lambda_a^{xw})$ and second-order topology in the following is similar.

In the region for $-2 + |b_{xw}| < M < -|b_{xw}|$ with $0 < |b_{xw}| < 1$, we have

$$C_{1m}^{xw}(\Lambda_1^{xw}) = 0, C_{1m}^{xw}(\Lambda_{2/3}^{xw}) = 1, C_{1m}^{xw}(\Lambda_4^{xw}) = -1. \tag{S30}$$

For $-4 + |b_{xw}| < M < -2 - |b_{xw}|$ with $0 < |b_{xw}| < 1$, we have

$$C_{1m}^{xw}(\Lambda_i^{xw}) = 0, C_{1m}^{xw}(\Lambda_4^{xw}) = 1, i = 1, 2, 3. \tag{S31}$$

In other regions, the system is a trivial insulator.

Note that for the region with non-zero $C_1$, the 4D parent system $\mathcal{H}_1$ is in the semimetal phase with gapless nodal-lines. For instance, when $M = 4$ and $0 < b_{xw} < 2$, there is a 4D nodal line centered at $\boldsymbol{k} = 0$, forming a circle at $k_x$-$k_w$ plane for $k_y = k_z = 0$. At four high-symmetric point, it hosts

$$C_1(\Lambda_1^{xw}) = 1, C_1(\Lambda_i^{xw}) = 0, i = 2, 3, 4. \tag{S32}$$

One can define a topological charge of this 4D nodal line as $Q = C_1^{in} - C_1^{out} = 1$ with $C_1^{in}$ is the first Chern number for $\mathcal{H}_1^{k_\perp}(k_\parallel)$ with $k_\perp = \sqrt{k_x^2 + k_w^2}$ insides the nodal line being $C_1(\Lambda_1^{xw})$, and $C_1^{out}$ is defined for the 2D subsystems outsides the nodal line being $C_1^{xw}(\Lambda_i)$. Similar cases can be discussed in other parameter regions with nodal-line structure, where we present this part in Sec. S-III C.

Note that one can calculate the corresponding MFCNs for $\mathcal{M}_{yz}$-symmetry and find $C_{1m}^{yz}(\Lambda_a^{yz}) = -C_{1m}^{xw}(\Lambda_a^{xw})$ accordingly. In addition, when only $b_{yw}$ is non-zero, one can easily follow our previous discussion and obtain the similar results accordingly.

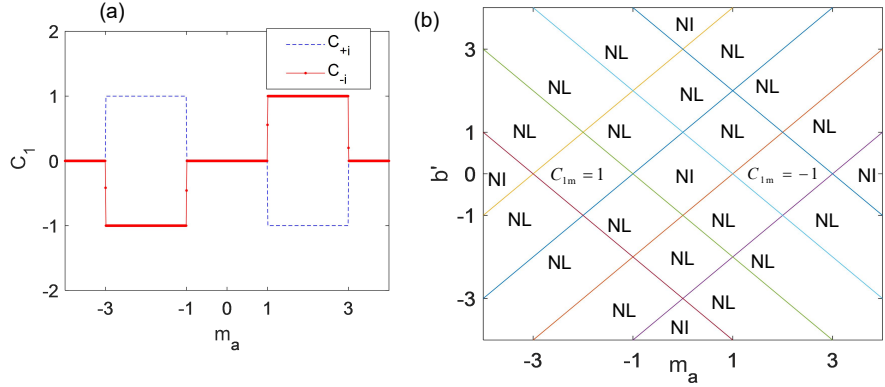

FIG. S2. (a) Numeric result of the phase diagram $(m_a, C_1)$ of $\mathcal{H}_\pm$ with $k_x = k_y$ and $b$ being zero. (b) The phase diagram of $\mathcal{H}_1$ with $k_x = k_y$ varying with $m_a$ and $b$. Here "NI" denotes normal insulators, "NL" denotes a 2D trivial nodal-line gapless structures. For a numeric small region of $b' = \sqrt{2}b$, we have a TI phase with non-trivial MFCN.

For $b_{xw} = b_{yw} = b$, we have $\mathcal{M}_{(x-y)z}$ symmetry with the eigenspace of $\hat{\mathcal{M}}_{(x-y)z}$ can be digonalized through

$$\begin{aligned}
U\hat{\mathcal{M}}_{(x-y)z}U^{-1} &= iG_{30}, \\
U &= \frac{1}{\sqrt{2}}\begin{pmatrix} 0 & e^{-i\frac{\pi}{4}} & 0 & -e^{-i\frac{\pi}{4}} \\ 0 & 1 & 0 & 1 \\ e^{-i\frac{\pi}{4}} & 0 & -e^{-i\frac{\pi}{4}} & 0 \\ 1 & 0 & 1 & 0 \end{pmatrix},
\end{aligned} \tag{S33}$$

where $U^{-1} = U^\dagger$. Applying this unitary transformation for $\mathcal{H}_1$ with $k_y = k_x$ and $k_z = 0, \pi$, we have

$$\begin{aligned}
U\mathcal{H}_1U^{-1} &= -\sqrt{2}\sin k_x G_{33} + \sqrt{2}bG_{30} - \sin k_w G_{02} \\
&\quad + (m_a - 2\cos k_x - \cos k_w)G_{01} = \mathcal{H}_+ \oplus \mathcal{H}_-,
\end{aligned} \tag{S34}$$

where

$$m_1 = M - 1, m_2 = M + 1 \tag{S35}$$

for $k_z = 0$ and $k_z = \pi$ respectively. The phase diagram of $\mathcal{H}_\pm$ characterized by $C_{\pm i}$ with $m_a$ and $b = 0$ is given by Fig. S2(a).

Here we emphasize that numeric small $b$ will not affect the value of $C_{\pm i}$ for each block $\mathcal{H}_\pm$ but it modifies the gapped regions of model (S34). The energy spectrum of (S34) is given by

$$E = \pm\sqrt{2\sin k_x^2 + \sin k_w^2 + (m_a - 2\cos k_x - \cos k_w)^2} \pm b'^2, \tag{S36}$$

leading to a modified phase diagram of the total system with the phase boundaries when the middle two-bands touching at

$$m_a \pm b' = \pm 1, \pm 3, \tag{S37}$$

i.e.,

$$b' = \pm m_a \pm 1, \pm m_a \pm 3 \tag{S38}$$

where $b' = \sqrt{2}b$. As shown in Fig. S2(b), in the gapped region for $b$ being numeric small, one can calculate the MFCN as we mentioned above with the result

$$C_{1m}^0 = -1, \ C_{1m}^\pi = 0, \tag{S39}$$

when $2 + \sqrt{2}|b| < M < 4 - \sqrt{2}|b|$, and

$$C_{1m}^0 = 0, \ C_{1m}^\pi = -1, \tag{S40}$$

when $\sqrt{2}|b| < M < 2 - \sqrt{2}|b|$, and

$$C_{1m}^0 = 1, \ C_{1m}^\pi = 0, \tag{S41}$$

when $-2 + \sqrt{2}|b| < M < -\sqrt{2}|b|$, and

$$C_{1m}^0 = 0, \ C_{1m}^\pi = 1, \tag{S42}$$

when $-4 + \sqrt{2}|b| < M < -2 - \sqrt{2}|b|$. In other regions $|M| > 4 - \sqrt{2}|b|$, the MFCNs are zero. With above results, we can characterize the second-order Fermi-arc in the 4D parent system with a non-zero $b$ accordingly.

Similarly, for $b_{xw} = -b_{yw} = b$, we have $\mathcal{M}_{(x+y)z}$ symmetry with the eigenspace of $\hat{\mathcal{M}}_{(x+y)z}$ can be digonalized through

$$
\begin{aligned}
U\hat{\mathcal{M}}_{(x+y)z}U^{-1} &= iG_{30}, \\
U &= \frac{1}{\sqrt{2}} \begin{pmatrix} 0 & -e^{i\frac{\pi}{4}} & 0 & e^{i\frac{\pi}{4}} \\ 0 & 1 & 0 & 1 \\ -e^{i\frac{\pi}{4}} & 0 & e^{i\frac{\pi}{4}} & 0 \\ 1 & 0 & 1 & 0 \end{pmatrix},
\end{aligned}
\tag{S43}
$$

where $U^{-1} = U^\dagger$. Applying this unitary transormation for $\mathcal{H}$ with $k_y = -k_x$ and $k_z = 0, \pi$, we have

$$
\begin{aligned}
U\mathcal{H}_1U^{-1} = &\sqrt{2}\sin k_x G_{33} - \sqrt{2}bG_{30} - \sin k_w G_{02} \\
&+ (m_\pm - 2\cos k_x - \cos k_w)G_{01},
\end{aligned}
\tag{S44}
$$

where $m_\pm = M \mp 1$ for $k_z = 0$ and $k_z = \pi$ respectively. One can calculate the MFCNs as we did above, which give opposite values of MFCNs at corresponding parameter regions of the case when $b_{xw} = b_{yz} = b$.

Note that the gapless region "NL" correponds to a gapless phase with 4D nodal-line structures in its 4D parent system when $|b|$ is large enough, similar to the case when only $b_{xw}$ or $b_{yz}$ is non-zero.

### 3. 3D chiral second-order topological subsystems with varying $k_w$

If we treat $k_w$ as a periodic parameter $\varphi$, 3D subsystems $\mathcal{H}_1^\varphi(k_x, k_y, k_z)$ with only $b_{xw} > 0$ being numeric small, host the combined mirror $\mathcal{M}_{yz}$ and together with the charge-conjugate,

$$\mathcal{C}\mathcal{H}_1^\varphi(k_x, k_y, k_z)\mathcal{C}^{-1} = -\mathcal{H}_1^\varphi(-k_x, -k_y, -k_z), \tag{S45}$$

with $\mathcal{C} = G_{22}\mathcal{K}$ satisfying $\mathcal{C}^2 = +1$. This model belongs to class D which is trivial in 3D according to AZ classification. Together with $\mathcal{M}_{yz}$ symmetry, the system indeed hosts second-order topology as we demonstrated below. By tuning the parameter $\varphi$, these 3D subsystems host second-order topology undergoing unconventional topological phase transitions. As shown in Fig. S3, 3D systems are in the nontrivial second-order topological phases when $\varphi \in (-\arcsin b_{xw}, 0) \cup (0, \arcsin b_{xw})$ with the phase transition points at $\varphi = 0, \pm \arcsin b_{xw}$ are topological crystalline insulators [7] where the surface Dirac cones close at $y = \pm L/2$, $x = \mp L/2$ planes, respectively. When $|\varphi| \geq \arcsin b_{xw}$, the system is trivial. Note that we have $\arcsin b_{xw} = b_{xw}$ when $b_{xw}$ being numeric small.

To characterize these second-order hinge states, we prefer to focus on the topology from its first-order boundaries at $x^\pm/y^\pm = \pm L/2$. Its surface Dirac Hamiltonian from Eq. (S23) are now given by

$$
\begin{aligned}
\mathcal{H}_{x,\pm}^D &= \mp k_y \sigma_2 \mp k_z \sigma_3 \pm m_{x^\pm} \sigma_1, \\
\mathcal{H}_{y,\pm}^D &= \mp k_x \sigma_1 \mp k_z \sigma_3 \pm m_{y^\pm} \sigma_2,
\end{aligned}
\tag{S46}
$$

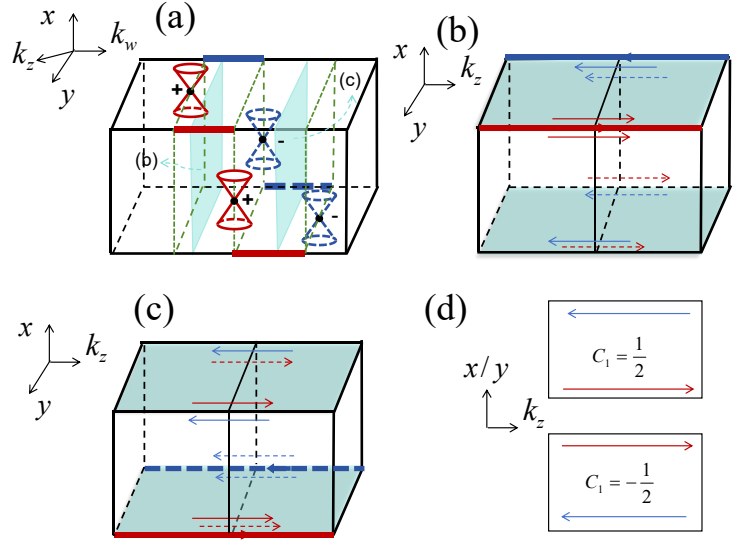

FIG. S3. (a) Schematic of the boundary modes of $\mathcal{H}_1$ with only $b_{xw}$ being numeric small and positive. We let $\varphi = k_w$ hereafter. (b) 3D SOTI phase when $\varphi \in (-\arcsin b_{xw}, 0)$. (c) 3D SOTI phase when $\varphi \in (0, \arcsin b_{xw})$. (d) Schematic of the chiral modes associated with the first Chern number at each surface.

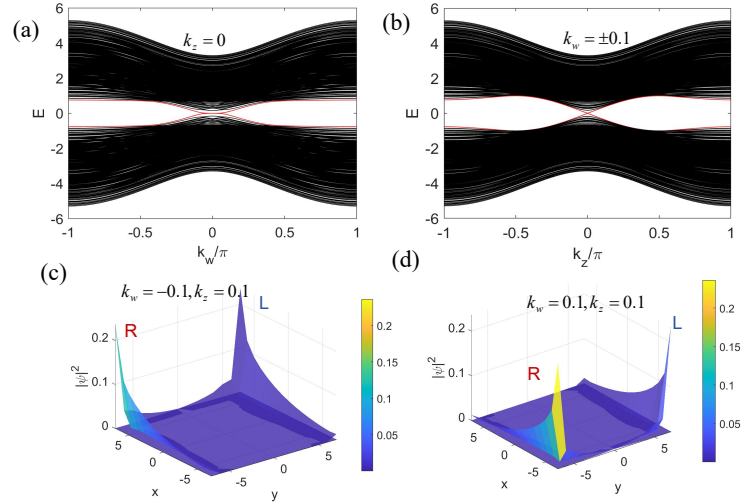

FIG. S4. Energy spectrum when considering OBC along $x$ and $y$ directions. (a)$E(k_w)$ when $k_z = 0$; (b) $E(k_z)$ when $k_w = \pm 0.1$. The location of chiral hinge modes:(c) when $k_w = -0.1$, and $k_z = 0.1$; (d) when $k_w = 0.1$, and $k_z = 0.1$. "L" and "R" denotes left and right chirality of the hinge modes respectively. In the above numeric, we set the lattice length $L = 15$, $M = 3$, and $b_{xw} = 0.1\pi$.

where $m_{x\pm} = \sin\varphi \pm b_{xw}$, and $m_{y\pm} = \sin\varphi$. Note that each massive Dirac cone at different surfaces contributes the first Chern number

$$C_1^x = \text{sign}(v_y v_z m_x)/2, C_1^y = \text{sign}(v_x v_z m_y)/2, \tag{S47}$$

with $v_{x,y,z} = \pm 1$, and $m_{x/y} = m_{x\pm/y\pm}$ at the corresponding 2D surfaces. For instance, in the region $\varphi \in (-\arcsin b_{xw}, 0)$ for $b_{xw} > 0$, we have

$$C_1^{x\pm} = 1/2, C_1^{y\pm} = \mp 1/2, \tag{S48}$$

and for $\varphi \in (0, \arcsin b_{xw})$,

$$C_1^{x\pm} = 1/2, C_1^{y\pm} = \pm 1/2. \tag{S49}$$

This leads to the non-vanishing chiral modes propagating along the two hinges at the top or buttom surfaces, as illustrated in Fig. S3(b-c). Here the propagation of the chiral modes related to the first Chern number for each surface is presented in Fig. S3(d). Similarly, one can easily check that the chiral modes are vanishing since $C_1^{x\pm} = \mp 1/2$ and $C_1^{y\pm} = \mp 1/2$ for $\varphi \in (-\pi, -\arcsin b_{xw})$; $C_1^{x\pm} = \pm 1/2$ and $C_1^{y\pm} = \pm 1/2$ for $\varphi \in (\arcsin b_{xw}, \pi)$. One can also solve the second-order hinge modes in analytical way by using the method developed in Ref. [5]. Here we present the numeric results that are consistent with our discussion above as in Fig. S4.

At the phase transition points $\varphi = 0, \pm \arcsin b_{xw}$, there are the crystalline insulators. For instance, when $\varphi = 0$, the system hosts surface Dirac points at the origin of the surface normal to $y$ and $z$ directions while the surface Dirac cones at the surfaces normal to $x$-direction are still gapped.

Similar situation appears when only $b_{yw}$ is non-zero. When both $b_{xw}$ and $b_{yw}$ survives, the system still hosts the second-order hinge states as the superposition of the above two cases unless the band gap closes and reopens.

## B.  $\mathcal{H}_1^{k_z=0}$ in Class AIII$^{\mathcal{M}_i}$

### 1.  Class AIII with $\mathcal{M}_{ij}/\mathcal{M}_i$ in 3D

By setting $k_z = 0$ in $\mathcal{H}_1$, we obtain a 3D novel TI phase $\mathcal{H}_1(k_x, k_y, k_w)$ restoring the chiral symmtery with $\hat{S} = \Gamma_3$, and thus falls into class AIII. The first-order boundary states become 2D Dirac cones characterized by the 3D winding number $w_3$ defined in Sec. S-II C while the second-order Fermi-arc states are now consisting of zero-energy modes along $k_w$.

In the case when only $b_{xw} \neq 0$, the system hosts both $\mathcal{M}_{xw}$ and $\mathcal{M}_y$ symmetries. We can define 1D winding number $w_1$ at four $\mathcal{M}_{xw}$-invariant points, and the mirror first Chern number at $\mathcal{M}_y$-invariant planes. Without loss of generality, we present a concrete discussion here. For $\mathcal{M}_{xw}$, we can also rotate $\hat{\mathcal{M}}_{xw}$ into $iG_{30}$ as we did in Sec. S-III A 2, and obtain the block diagonal Hamiltonian at high-symmetry points as Eq. (S50) by setting $k_z = 0$, given by,

$$U_{xw}\mathcal{H}_1^{k_z=0}(\Lambda_a^{xw})U_{xw}^{-1} = \sin k_y G_{32} - (m_a - \cos k_y)G_{31} + b_{xw}G_{01} = \mathcal{H}_+ \oplus \mathcal{H}_-, \tag{S50}$$

where $m_a = m - \cos k_x - \cos k_w$ at $\Lambda_a^{xw}$ being

$$m_1 = m - 2, m_2 = m, m_3 = m, m_4 = m + 2, \tag{S51}$$

with $m = M - 1$. At each $\Lambda_a^{xw}$, we can define the 1D winding number $w_{\pm i}$ for $\mathcal{H}_\pm$ as presented in Sec. S-II A. As we can see that the phase boundaries of $\mathcal{H}_\pm$ for $b_{xw} = 0$ is $m_a = \pm 1$, then we have new phase boundaries as

$$m_a \pm b_{xw} = \pm 1 \rightarrow b_{xw} = \pm m_a \pm 1. \tag{S52}$$

Note that the sum of the winding number $w_1 = w_{+i} + w_{-i}$ is non-zero since this 1D system in class AIII belongs to $\mathbb{Z}$ class. The phase diagram of this system is shown in Fig. S5.

Similarly, we have

$$w_1(\Lambda_1^{xw}) = 2, w_1(\Lambda_i^{xw}) = 0, i = 2, 3, 4 \tag{S53}$$

when $1 + |b_{xw}| < m < 3 - |b_{xw}|$ and $0 < |b_{xw}| < 1$. This result implies that there exists second-order Fermi-arcs located at the position with mirror reflection about $k_w = 0$ and $y = 0$ hinges, similar to the case in Fig. S3(a).

For $\mathcal{M}_y$-symmetry, we rotate $\hat{\mathcal{M}}_y = iG_{01}$ into $iG_{30}$ with $U_y = e^{-i\frac{\pi}{4}G_{13}}e^{i\frac{\pi}{4}G_{22}}$ in the representation as

$$U_y\mathcal{H}_1^{k_y=0,\pi}U_y^{-1} = \sin k_x G_{31} + \sin k_w G_{33} + (m_a - \cos k_x - \cos k_w)G_{32} + b_{xw}G_{30}, \tag{S54}$$

where $m_1 = m - 1$ and $m_2 = m + 1$ for $k_y = 0$ and $k_y = \pi$ respectively. Now the phase boundaries determined at the position when the bulk gap closes at $E = 0$ are modified as

$$m_a \pm b_{xw} = 0, \pm 2 \rightarrow b_{xw} = \pm m_a, \pm m_a \pm 2. \tag{S55}$$

The system actually host the exact same phase diagram as in Fig. S1. One can also obtain the topological number of this system as

$$C_{1m}^0 = -1, C_{1m}^\pi = 0, \tag{S56}$$

when $1 + |b_{xw}| < m < 3 - b_{xw}$. This results describe that the second-order Fermi-arcs are located at $\mathcal{M}_y$ reflection hinges about $y = 0$ plane.

In the region when $w_1 = 1$ and $C_1$ is non-zero, the original 3D system $\mathcal{H}_1^{k_z=0}(k_x, k_y, k_w)$ is in the nodal-line semimetallic phase.

For $m$ is in other parameter regions and the case for $b_{yw}$ is non-zero, there are similar discussions based on our previous spirit can be recovered. In addition, $\mathcal{M}_{x \pm y}$-symmetry still hosts when $b_{xw} = \pm b_{yw}$ where the MFCN can be also defined at $k_x = \pm k_y$ plane. We leave this part for the interested readers.

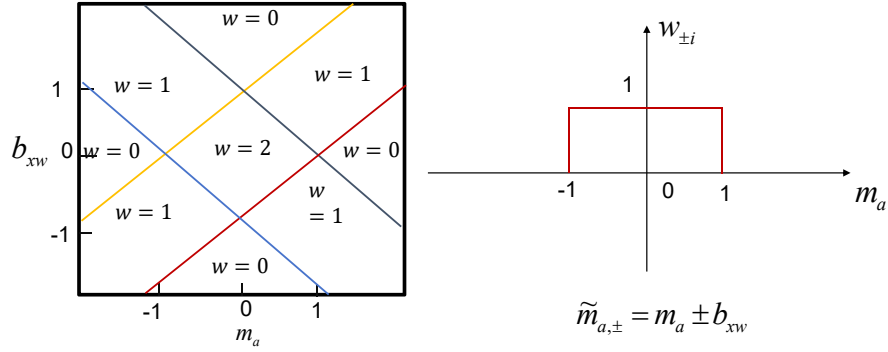

FIG. S5. Left panel: Schematic of the phase diagram of the effective 1D system $\mathcal{H}_1^{k_z=0}(\Lambda_a^{xw})$ with parameters $(m_a, b_{xw})$. The straight color lines denote the phase transition boundaries where the bulk gap closes at $E = 0$. Note that $w_1 = w_{+i} + w_{-i}$. Right panel: The schematic of the phase diagram of $\mathcal{H}_\pm$ as the conventional 1D topological insulators with $w_{\pm i}$ by varying $m_a$ when $b_{xw} = 0$.

### 2. 2D second-order topological subsystems with varying $k_w$

If we treat $k_w$ as a periodic parameter $\varphi$, 3D subsystems $\mathcal{H}_1^{\varphi, k_z=0}(k_x, k_y)$ with only $b_{xw}$ being numeric small go through unconventional topological phase transitions when tuning $\varphi$. It now hosts $\mathcal{T}$, $\mathcal{C}$ and $\mathcal{S}$-symmetries with $\hat{\mathcal{T}} = G_{11}\mathcal{K}$, $\hat{\mathcal{C}} = G_{22}\mathcal{K}$, and $\hat{\mathcal{S}} = G_{33}$, and thus falls into class BDI which is trivial in 2D. Together with $\mathcal{M}_y$, it becomes a non-trivial topological crystalline insulator. Its 1D edge Dirac Hamiltonian from Eq. (S23) are now given by

$$
\begin{aligned}
\mathcal{H}_{x,\pm}^D &= \mp k_y \sigma_2 \pm m_{x\pm} \sigma_1, \\
\mathcal{H}_{y,\pm}^D &= \mp k_x \sigma_1 \pm m_{y\pm} \sigma_2,
\end{aligned}
\tag{S57}
$$

where $m_{x\pm} = \sin\varphi \pm b_{xw}$, and $m_{y\pm} = \sin\varphi$. Note that each massive Dirac cone at different edges contributes the half integer of winding number. The location of the corner charges can be analytically determined by sloving the differential equations of each edges as usual, we refer to Ref. [5] for details. The numeric results are similar to the case in Fig. S4(a,c,d) when taking $k_z = 0$. At three phase transition points $\varphi = 0, \pm\arcsin b_{xw}$, the systems are 2D topological insulators with gapped Dirac modes on the selective edges. See Fig. S6 for the whole picture of this phase.

As similar case happens when only $b_{yw}$ is non-zero. The case when $b_{xw}$ and $b_{yw}$ are non-zero can be also recovered akin to our previous discussions. We leave this part for the interested readers.

### C. Nodal-line semimetal phases and field-theoretical descriptions

For a large value of $b_{xw}/b_{yw}$, the bulk gap of $\mathcal{H}_1$ will close and become a nodal-line semimetal phase. Without loss of generality and for concreteness, we focus on the case when $M$ near 4 and $b_{xw} > 0$ being small, one can expand $\mathcal{H}_1$ near the origin of BZ, then we have

$$
\mathcal{H}_{eff} = k_x\Gamma_1 + k_y\Gamma_2 + k_z\Gamma_3 + k_w\Gamma_4 + m_0\Gamma_0 + b_{xw}\Gamma_0\Gamma_{14}
\tag{S58}
$$

with the energy spectrum

$$
E = \pm\sqrt{k_y^2 + k_z^2 + (\sqrt{k_x^2 + k_w^2 + m_0^2} \pm b_{xw})^2}
\tag{S59}
$$

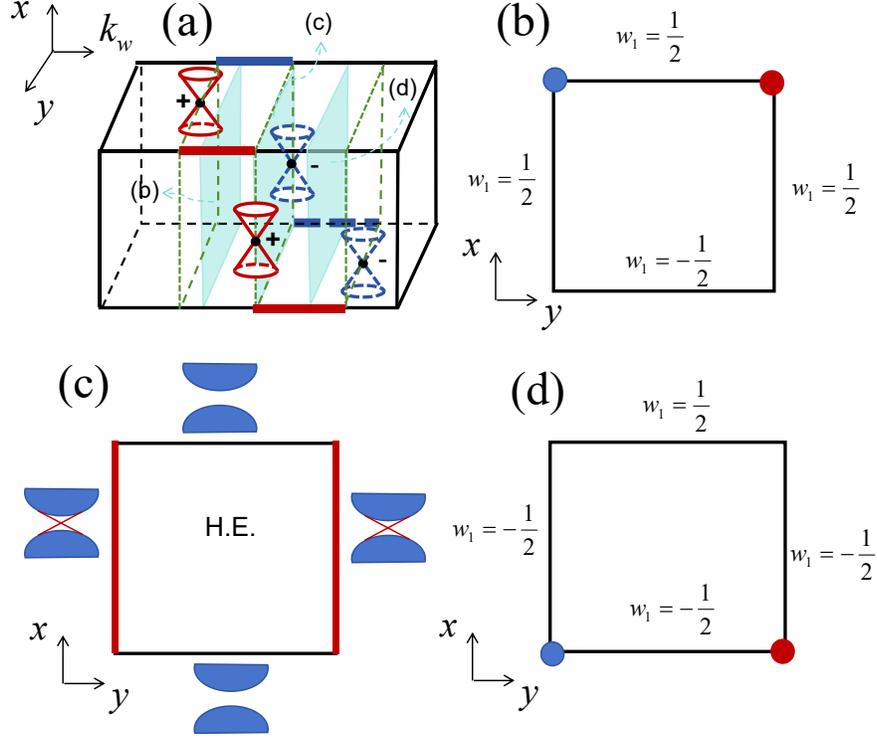

FIG. S6. (a) Schematic of the boundary modes of $\mathcal{H}_1^{k_z=0}$ with only $b_{xw}$ being numeric small and positive. 2D surface Dirac points colored in red and blue host right and left vorticity respectively. We let $\varphi = k_w$ hereafter. (b) 2D SOTI phase with two corner states when $\varphi \in (-\arcsin b_{xw}, 0)$. (c)Topological crystalline insulator hosts helical edge states at $y = \pm L/2$ at phase transition point $\varphi = 0$. (d) 2D SOTI phase with two corner states when $\varphi \in (0, \arcsin b_{xw})$.

where $m_0 = m + \frac{1}{2}(k_y^2 + k_z^2)$ with $m = M - 4$. As we see that there is a nodal-line structure of band crossing points satisfying the equation:

$$k_x^2 + k_w^2 = \rho^2, \text{ with } \rho = \sqrt{b_{xw}^2 - m^2} \tag{S60}$$

at $k_y = k_z = 0$ plane when $|b_{xw}| > |m|$. Note that this nodal-line lives on $\mathcal{M}_{yz}$-invariant plane can be characterized by the first Chern number defined in the 2D subsystems inside and outside the nodal-line as we mentioned in Sec. S-III A 2. Thus this nodal structure can be treated as stacking of 2D Chern insulators in $k_x$-$k_w$ plane with a band crossing region forming a nodal-loop seperates the non-trivial and trivial phases inside and outside the nodal line respectively. Alternately, this nodal line can be also seen as the stacking of a pair of Weyl point dipole at fixed angle $\theta$ when running over $\theta \in [0, 2\pi]$.

Note that the effective model Hamiltonian of the two middle-bands can be obtained through the first-order perturbation theory[6], which is given by,

$$H_{NL}^{4D} = k_z\sigma_1 - k_y\sigma_2 + (b_{xw} - \sqrt{k_x^2 + k_w^2 + m_0^2})\sigma_3. \tag{S61}$$

Next we focus on the topological field thereotical description of this nodal structure. The effective action of this nodal line structure is given by

$$S_{NL} = \frac{\mathcal{B}_{ab}}{16\pi^3} \int e^a \wedge e^b \wedge A \wedge dA, \tag{S62}$$

where $e^a = e^a_\mu dx^\mu$ denotes the translation/elastic gauge field[8, 9].

To obtain this effective action, we can first use the $LU(1)$ theory [8, 10]. We start by considering the $d = (4+1) + 1 + 1 = 7$ dimensional Chern-Simons theory [11],

$$S = \frac{1}{4!(2\pi)^3} \int \mathcal{A} \wedge d\mathcal{A} \wedge d\mathcal{A} \wedge d\mathcal{A}. \tag{S63}$$

Inserting $\mathcal{A} = A + xK_x + wK_w$ into the above action, we have

$$S = \frac{1}{4!(2\pi)^3}\int\ (A + xK_x + wK_w)\wedge d(A + xK_x + wK_w)\wedge d(A + xK_x + wK_w)\wedge d(A + xK_x + wK_w). \quad (S64)$$

Here $A$ denotes the $U(1)$ gauge field, $x,w$ denote the translation gauge fields and $K_{x,w}(\theta)$ are the coordinates of the points on the nodal line ($\theta \in [0, 2\pi]$) with the radius $\rho$ when only $b_{xw}$ is non-zero. This gives rise to a mixed anomaly term, except in one extra dimension

$$S = 6 \times \frac{1}{4!(2\pi)^3}\int\ [d(xK_x + wK_w)\wedge d(xK_x + wK_w)\wedge A\wedge dA + (xK_x + wK_w)\wedge d(xK_x + wK_w)\wedge dA\wedge dA], \quad (S65)$$

corresponding to a boundary theory (note that $d^2 = 0$)

$$\begin{aligned}
S &= \frac{1}{4(2\pi)^3}\int\ (xK_x + wK_w)\wedge d(xK_x + wK_w)\wedge A\wedge dA\\
&= \frac{1}{4(2\pi)^3}\int\ (K_x\partial_\theta K_w - K_w\partial_\theta K_x)x\wedge w\wedge A\wedge dA\\
&= \frac{V_F}{16\pi^3}\int\ x\wedge w\wedge A\wedge dA
\end{aligned} \quad (S66)$$

where $V_F = \frac{1}{2}\int d\theta(K_x\partial_\theta K_w - K_w\partial_\theta K_x) = \pi\rho^2$ is the area in the momentum space enclosed by the nodal line if we parameterize $K_x = \rho\cos\theta$ and $K_w = \rho\sin\theta$.

By replacing $x \to e^x$, $w \to e^w$, and $V_F \to \mathcal{B}_{xw}$, we obtain the action as presented in Eq. (S62). The time-component term comes from $\mathcal{A} = A + xK_x + wK_w + tK_t$, where the nodal line now tilts along $k_x$ direction in energy $E$ with $K_t = \sqrt{\rho^2 - K_w^2}\sin\varphi$ and $\cos\varphi = K_x/\sqrt{\rho^2 - K_w^2}$. Then we obtain

$$\mathcal{B}_{tw} = (K_t\partial_\theta K_w - K_w\partial_\theta K_t) \quad (S67)$$

which is nothing but the projection area enclosed by the nodal line at $k_w$-$E$ plane.

For $e_x^x = 1$, $e_w^w = 1$, i.e., without lattice deformation, we have

$$S_{NL}^{4D} = \frac{\mathcal{B}_{xw}}{16\pi^3}\int d^5 x\epsilon^{xw\mu\nu\lambda}A_\mu\partial_\nu A_\lambda + \frac{\mathcal{B}_{tw}}{16\pi^3}\int d^5 x\epsilon^{tw\mu\nu\lambda}A_\mu\partial_\nu A_\lambda. \quad (S68)$$

As we can see that the first term is nothing but the summation of $(2+1)$D Chern-Simons action for each 2D Chern insulator $\mathcal{H}_1^{k_x - k_w}(k_y, k_z)$ running over $k_x$-$k_w$ plane. $\mathcal{B}_{xw}$ denotes the area enclosing the nodal line projected onto the boundary BZ corresponding to the nontrivial Chern insulators contributes the first Chern number $C_1 = 1$ while the 2D insulators are trivial with $C_1 = 0$ in the outside region, as depicted in Fig. S7(a). If we introduce a perturbation as $\Delta_x = \delta_x\sin k_x$ into $\mathcal{H}_1$, it tilts the nodal line along $k_x$-direction in energy. The second term in Eq. (S68) is used to describe this case which can be treated as the stacking of the action arising from chiral anomaly of two tilted Weyl points [12–14] along $k_w$ direction, as shown in Fig. S7(b). The corresponding topological currents can be obtained by varying $S_{NL}^{4D}$ with $A_\mu$, i.e., $J^\mu = \frac{\delta S_{NL}^{4D}}{\delta A_\mu}$. If we pick a gauge field as $A_\mu = (0, zB_y, 0, tE_z, 0)$, we have

$$J^y = \frac{\delta S_{NL}^{4D}}{\delta A_y} = \frac{\mathcal{B}_{xw}}{8\pi^3}E_z + \frac{\mathcal{B}_{tw}}{8\pi^3}B_y, \quad (S69)$$

where the first term denotes the summation of the quantum Hall currents in $k_x$-$k_w$ plane enclosed the nodal-line while the second-term is the summation of the chiral magnetic currents [12–14] between two tilted Weyl points with fixed $(\theta_0, \varphi_0)$ in energy when running over the whole $(\theta, \varphi)$ space. By simply setting $k_z = 0$, we obtain a 3D nodal line with the effective Hamiltonian as we mentioned above. The effective action of this nodal line has been investigated in recent work[8, 9, 15] which can be derived in the same spirit as

$$S_{NL}^{3D} = \frac{\mathcal{B}_{xw}}{8\pi^2}\int d^4 x\epsilon^{xw\mu\nu}\partial_\mu A_\nu + \frac{\mathcal{B}_{tw}}{8\pi^2}\int d^4 x\epsilon^{tw\mu\nu}\partial_\mu A_\nu \quad (S70)$$

where $x^\mu = (t, x, y, w)$.

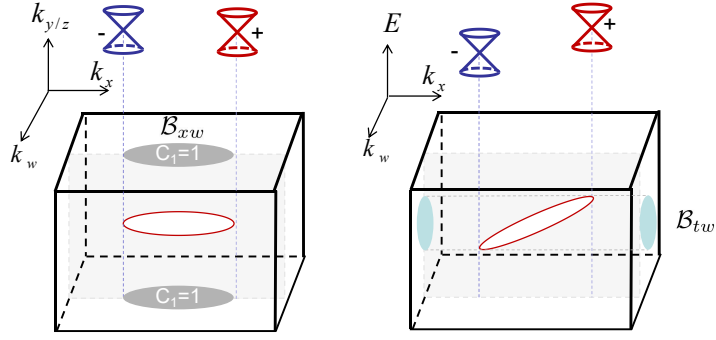

FIG. S7. Schematic of 4D nodal line which can be treated as the stacking of 2D Chern insualtors along $k_x$-$k_w$ direction or a pair of 3D Weyl points along $k_w$ axis:(a)for $\delta_x = 0$, the projection area of nodal line at the boundary BZ is $\mathcal{B}_{xw}$ in $k_x$-$k_w$ plane; (b)for $\delta_x \neq 0$, nodal line tilts along $k_x$ direction in energy with the projection area $\mathcal{B}_{tw}$ in $E$-$k_w$ plane.

## S-IV. $\mathcal{H}_2$ IN CLASS AIII$^{\mathcal{RT}}$ WITH AND WITHOUT SPIN-ORBIT-COUPLING AND $\mathcal{H}_2^{k_z=0}$ IN CLASS AIII WITH SPIN CONSERVATION

### A. $\mathcal{H}_2$ in class AIII$^{\mathcal{RT}}$ with and without spin-orbit-coupling

#### 1. Shifted boundary Dirac nodal structures

Consider the eight-band model $\mathcal{H}_2 = \tilde{\mathcal{H}}_0 + \Delta_2 + \Delta_3$ as presented in the main text, where

$$\tilde{\mathcal{H}}_0 = \sin k_x \tilde{\Gamma}_1 + \sin k_y \tilde{\Gamma}_2 + \sin k_z \tilde{\Gamma}_3 + \sin k_w \tilde{\Gamma}_4 + (M - \sum_{i=x,y,z,w} \cos k_i) \tilde{\Gamma}_0, \tag{S71}$$

and

$$\begin{aligned}
\Delta_2 &= b_{xw} \tilde{\Gamma}_0 \tilde{\Gamma}_{14} + b_{yw} \tilde{\Gamma}_0 \tilde{\Gamma}_{24} + b_{zw} \tilde{\Gamma}_0 \tilde{\Gamma}_{34}, \\
\Delta_3 &= c_x \tilde{\Gamma}_{61} + c_z \tilde{\Gamma}_{63}.
\end{aligned} \tag{S72}$$

The projectors upon applying open boundary condition along $\alpha$-direction are given by [6]

$$P_\alpha^\pm = \frac{1}{2}(1 \mp i\tilde{\Gamma}_\alpha \tilde{\Gamma}_0), \tag{S73}$$

where $\pm$ represents the projection of $\mathcal{H}_2$ to the boundary at $\alpha = \pm L/2$, with a lattice length of $L$. Here we mix the notation in the subscript of $\tilde{\Gamma}_\alpha$: $\alpha = (x, y, z, w) \leftrightarrow (1, 2, 3, 4)$.

Thus the effective Hamiltonian at the first-order boundaries for $\mathcal{H}_2$ is given by,

$$\mathcal{H}_{\alpha,eff}^\pm = P_\alpha^\pm \mathcal{H}_2 P_\alpha^\pm, \tag{S74}$$

for $|\lambda_\alpha| < 1$ where

$$\lambda_\alpha = M - \sum_{i,i \neq \alpha} \cos k_i. \tag{S75}$$

In the diagonal representation of the projectors, i.e., $u_\alpha P_\alpha^+ u_\alpha^{-1} = 1_2 \oplus 0_2 \oplus 1_2 \oplus 0_2$, and $u_\alpha P_\alpha^- u_\alpha^{-1} = 0_2 \oplus 1_2 \oplus 0_2 \oplus 1_2$ with $1_2$ denotes the $2 \times 2$ identity matrix and $0_2$ is zero matrix in a $2 \times 2$ representation, we have all boundary effective Hamiltonian for the non-zero block, given by,

$$\begin{aligned}
\mathcal{H}_{x,eff}^\pm &= \mp \sin k_y G_{32} \mp \sin k_z G_{33} \pm (\sin k_w \pm b_{xw}) G_{31} \mp c_z G_{23}, \\
\mathcal{H}_{y,eff}^\pm &= \mp \sin k_x G_{31} \mp \sin k_z G_{33} \pm (\sin k_w \pm b_{yw}) G_{32} \mp c_x G_{21} \mp c_z G_{23}, \\
\mathcal{H}_{z,eff}^\pm &= \mp \sin k_x G_{31} \mp \sin k_y G_{32} \pm (\sin k_w \pm b_{zw}) G_{33} \mp c_x G_{21}, \\
\mathcal{H}_{w,eff}^\pm &= \mp (\sin k_x \mp b_{xw}) G_{31} \mp (\sin k_y \mp b_{yw}) G_{32} \mp (\sin k_z \mp b_{zw}) G_{33} \mp c_x G_{21} \mp c_z G_{23},
\end{aligned} \tag{S76}$$

with $u_x = e^{i\frac{\pi}{4}G_{011}}$, $u_y = e^{i\frac{\pi}{4}G_{012}}$, $u_z = e^{i\frac{\pi}{4}G_{013}}$, and $u_w = 1_8$ being the identity.

For $c_x = c_z = 0$, there is a shifted Dirac cone at $x/y/z = \pm L/2$ along $k_w$ axis when $2 < M < 4$, whose low-energy Hamiltonian near the origin is given by

$$\begin{aligned}
\mathcal{H}_{x,\pm} &= \mp k_y G_{32} \mp k_z G_{33} \pm (k_w \pm b_{xw}) G_{31}, \\
\mathcal{H}_{y,\pm} &= \mp k_x G_{31} \mp k_z G_{33} \pm (k_w \pm b_{yw}) G_{32}, \\
\mathcal{H}_{z,\pm} &= \mp k_x G_{31} \mp k_y G_{32} \pm (k_w \pm b_{zw}) G_{33},
\end{aligned} \tag{S77}$$

respectively. At the boundary $x = \pm L/2$, there is a Dirac cone located at $\mathbf{k}_{x\pm} = (k_y, k_z, k_w) = (0, 0, \mp b_{xw})$ with chirality $\chi_\pm = \pm 1$; At $y = \pm L/2$, there is a Dirac cone located at $\mathbf{k}_{y\pm} = (k_x, k_z, k_w) = (0, 0, \mp b_{yw})$ with chirality $\chi_\pm = \pm 1$; At $z = \pm L/2$, there is a Dirac cone located at $\mathbf{k}_{z\pm} = (k_x, k_y, k_w) = (0, 0, \mp b_{zw})$ with chirality $\chi_\pm = \pm 1$. Note that $b_{xw}/b_{yw}/b_{zw}$ both acts on the boundaries normal to $x/y/z$ and $w$ direction, shifting the Weyl cones along $k_w$ and $k_x/k_y/k_z$ directions, respectively.

For $b_{xw} = b_{yw} = b_{zw} = 0$, each Dirac cone is expanded into a 3D nodal line with the Hamiltonian

$$\begin{aligned}
\mathcal{H}_{x,\pm} &= \mp k_y G_{32} \mp k_z G_{33} \pm k_w G_{31} \mp c_z G_{23}, \\
\mathcal{H}_{y,\pm} &= \mp k_x G_{31} \mp k_z G_{33} \pm k_w G_{32} \mp c_x G_{21} \mp c_z G_{23}, \\
\mathcal{H}_{z,\pm} &= \mp k_x G_{31} \mp k_y G_{32} \pm k_w G_{33} \mp c_x G_{21}.
\end{aligned} \tag{S78}$$

For instance, $\mathcal{H}_{x,+}$ describes a 3D nodal line sturcture with the energy spectrum

$$E = \pm\sqrt{k_z^2 + (\sqrt{k_y^2 + k_w^2} + c_z)^2}, \tag{S79}$$

where its band crossing at $E = 0$ forming a nodal line at $k_z = 0$ plane with the equation

$$k_y^2 + k_w^2 = c_z^2. \tag{S80}$$

Note that this is a second-order nodal line protected by the $\mathcal{IT}$-symmetry with $(\mathcal{IT})^2 = +1$ carrying double topological charges [5, 16].

When tuning on all the parameters $b_{xw}$, $b_{yw}$, $b_{zw}$, and $c_{x,z}$, each first-order boundary hosts shifted 3D Dirac nodal line states instead.

### 2. Calculation of spin mirror first Chern number and mirror winding number

For $\mathcal{H}_2$ with $c_x = c_z = 0$, the system is spin conservation and thus hosts $\tilde{\Gamma}_5$-symmetry. For its first-order topology, one can define the spin second Chern number (SSCN) as the half difference of $C_2^\pm$ in two spin block as $C_{2s} = (C_2^+ - C_2^-)/2$.

To characterize its second-order topology, we define the so-called spin mirror first Chern number (SMFCN) at the mirror-invariant points. Without loss of generality, we just consider the case when only $b_{xw}$ is non-zero, and so on. The system $\mathcal{H}_2$ now is the two copies of $\pm\mathcal{H}_1$ with opposite MFCNs under $\mathcal{M}_{xw}$ and $\mathcal{M}_{yz}$ symmetries, i.e., $C_{1m,+}^{ij}(\Lambda) + C_{1m,-}^{ij}(\Lambda) = 0$ with $ij = xw, yz$, and thus we define SMFCN at $\Lambda_a^{ij}$ as

$$C_{1sm}^{ij}(\Lambda_a^{ij}) = \left(C_{1m,+}^{ij}(\Lambda_a^{ij}) - C_{1m,-}^{ij}(\Lambda_a^{ij})\right)/2, \tag{S81}$$

which takes the same value of $C_{1m,+}^{ij}(\Lambda_a^{ij})$ as we have evaluated in Sec. S-III A 2.

Similarly, for the case when $b_{xw} = b_{yw} = b_{zw} = b$, the system breaks $\mathcal{M}_{ij}$ but preserves $\mathcal{M}_{x-y}$, $\mathcal{M}_{y-z}$, and $\mathcal{M}_{z-x}$ symmetries determining the location of second-order Fermi-arcs. One can also calculate the 3D mirror winding numbers at the mirror invariant plane. For instance, we consider to rotate $\hat{\mathcal{M}}_{x-y} = \frac{1}{\sqrt{2}}(\tilde{\Gamma}_1 - \tilde{\Gamma}_2)\tilde{\Gamma}_7 = \frac{i}{\sqrt{2}}(G_{132} - G_{131})$ into $iG_{003}$, the Hamiltonian with $k_y = k_x$ is given by,

$$U\mathcal{H}_2 U^{-1} = -\sqrt{2}\sin k_x G_{200} + \sin k_z G_{333} + \sin k_w G_{320} + (M - 2\cos k_x - \cos k_z - \cos k_w)G_{310} - \sqrt{2}b G_{230} + b G_{303}, \tag{S82}$$

where we take $U = e^{-i\frac{\pi}{4}\frac{1}{\sqrt{2}}(G_{132} + G_{131})}$. The corresponding block Hamiltonian in the eigenspace $\mathcal{E} = \pm i$ of $\hat{\mathcal{M}}_{x-y}$ is

$$\mathcal{H}_\pm = -\sqrt{2}\sin k_x G_{20} \pm \sin k_z G_{33} + \sin k_w G_{32} + (M - 2\cos k_x - \cos k_z - \cos k_w)G_{31} - \sqrt{2}b G_{23} \pm b G_{30}. \tag{S83}$$

For $b = 0$, $\mathcal{H}_\pm$ hosts opposite 3D winding numbers [see the results calculated from Eq. (S15) in Fig. S9(a)] and its nature remains host when we tune on $b$ being numeric small. Thus we define a mirror winding number (when its 4D parent system is still gap) as

$$w_{3m} = (w_3^{+i} - w_3^{-i})/2. \tag{S84}$$

When $M = 3$ and $b = 0.1$, we have $w_3^{+i} = -w_3^{-i} = 1$ resulting in $w_{3m} = 1$. It implies the existence of helical gapless modes popagating along the $\mathcal{M}_{x-y}$-symmetric hinges about the $x = y$ plane. The similar story related to $\mathcal{M}_{y-z}$ and $\mathcal{M}_{z-x}$-symmetries can be repeated accordingly. The second-order helical states propagating along the corresponding hinges under OBCs along $x$, $y$, and $z$ directions are presented in Fig. S8(a).

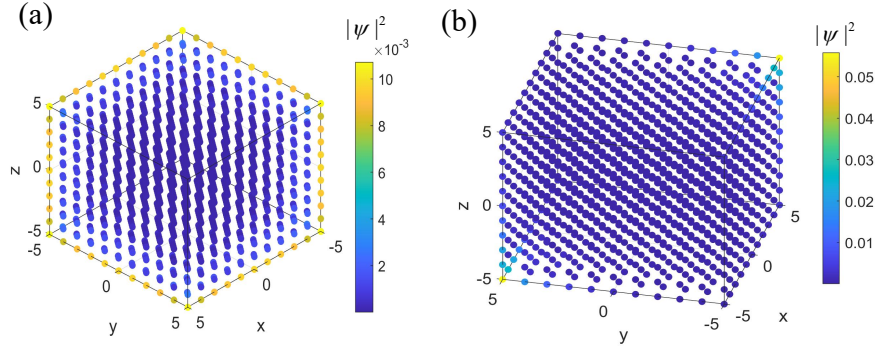

FIG. S8. (a)The state density distribution $|\psi|^2$ (colored in yellow) of the second-order hinge states for $\mathcal{H}_2$ at the slice $k_w = 0$, and paramters $M = 3$, $b = 0.3$, and $c = 0$. (b) The state density distribution $|\psi|^2$ of two third-order zero-energy corner states (colored in yellow) for $\mathcal{H}_2$ at the slice $k_w = 0$, and parameters $M = 3$, $b = 0.3$, and $c = 0.2$. Above we consider OBCs along $x$, $y$, and $z$ directions with the chain length $L = 11$.

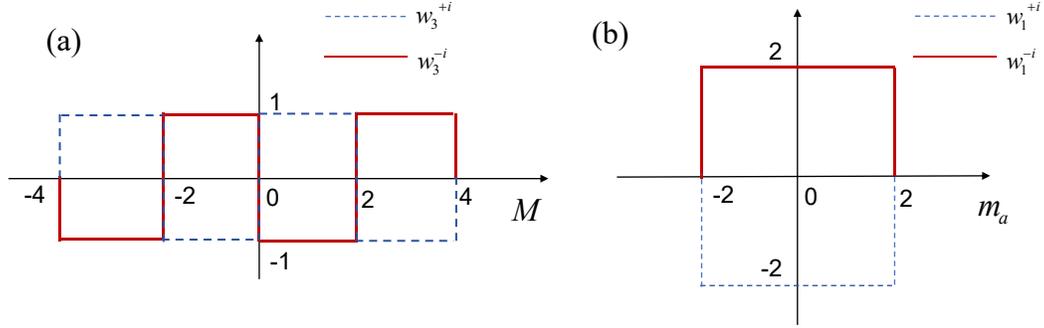

FIG. S9. (a)The schematic of the phase diagram of $\mathcal{H}_\pm$ in Eq. (S83) as the conventional 3D topological insulators with $w_3^{\pm i}$ by varying $M$ when $b = 0$. The phase transition points when bulk gap closes are $M = 0, \pm 2, \pm 4$. (b) The schematic of the phase diagram of $\mathcal{H}_\pm$ in Eq. (S86) as the conventional 1D topological insulators with $w_1^{\pm i}$ by varying $m_a$ when $b = c = 0$. The phase transition points when bulk gap closes are $m_a = \pm 2$.

When tuning $c_x = c_z = c$, i.e., we introduce the spin-orbit-coupling term $\Delta_3$, the first-order boundary modes become a Dirac nodal line structure. Spin SCN is not well-defined, one can define a $\mathbb{Z}_2$ number $\nu_{k_w}$ at the $\mathcal{RT}_w$-invariant planes at $k_w = 0, \pi$ to describe the first-order topology. In this case, the 3D subsystems with $m_\pm = M \mp 1$ are the so-called real topological insulators [17] or Takagi insulators preserving chiral and spinless $\mathcal{IT}$-symmetries, which have been studied in Ref. [18]. For instance, we find $\nu_0 = 1$ and $\nu_\pi = 0$ when $M = 3$, $b = 0.2$, and $c_x = c_z = 0.1$. Say, numeric small $\Delta_2$ and $\Delta_3$ will not change $\nu_{k_w}$ of $\tilde{\mathcal{H}}_0^{k_w = 0, \pi}$.

To characterize the third-order topology, we focus on $\mathcal{M}_{(x+z)yw}$-symmetry with the operator $\hat{\mathcal{M}}_{(x+z)yw} = \hat{\mathcal{M}}_{z+x}\hat{\mathcal{M}}_y\hat{\mathcal{M}}_w = \frac{i}{\sqrt{2}}(G_{223} - G_{221})$. One can define 1D mirror winding number at the mirror-invariant planes $k_z = -k_x$, and $\Lambda_a^{yw} = (k_y, k_w) = (0, 0), (0, \pi), (\pi, 0), (\pi, \pi)$. By rotating $\hat{\mathcal{M}}_{(x+z)yw}$ into $iG_{003}$, we have

$$U\mathcal{H}_2^{k_z = -k_x}(\Lambda_a^{yw})U^{-1} = \sqrt{2}\sin k_x G_{113} + (m_a - 2\cos k_x)G_{310} + \sqrt{2}bG_{120} - bG_{303} + \sqrt{2}cG_{010}, \tag{S85}$$

with $U = e^{i\frac{\pi}{4}G_{001}}e^{-i\frac{\pi}{4}\frac{1}{\sqrt{2}}(G_{223}+G_{221})}$. The corresponding block Hamiltonian in the eigenspace $\mathcal{E} = \pm i$ of $\hat{\mathcal{M}}_{(z+x)yw}$ is

$$\mathcal{H}_{\pm}(\Lambda_a^{yw}) = \pm\sqrt{2}\sin k_x G_{11} + (m_a - 2\cos k_x)G_{31} + \sqrt{2}bG_{12} \mp bG_{30} + \sqrt{2}cG_{01}, \tag{S86}$$

where $m_1 = M-2$, $m_2 = M$, $m_3 = M$, and $m_4 = M+2$. For $b = c = 0$, $\mathcal{H}_{\pm}$ hosts 1D winding number $w_{\pm i} = \pm 2$ for $-2 < m_a < 2$ and $w_{\pm i} = 0$ for elsewhere [see the results calculated from Eq. (S11) in Fig. S9(b)]. For numeric small $b$ and $c$, $w_1^{\pm i}$ will not changed but the phase boundaries will be modified. For example, we focus on the insulating phase when $M = 3$, $b = 0.2$ and $c = 0.1$. Since $w_1^{+i} + w_1^{-i} = 0$, we define 1D mirror winding number as

$$w_{1m} = (w_1^{+i} - w_1^{-i})/2, \tag{S87}$$

then we have

$$w_{1m}(\Lambda_1^{yw}) = 2, w_{1m}(\Lambda_i^{yw}) = 0, i = 2,3,4. \tag{S88}$$

This results guantees the exact location of the third-order Fermi-arc, which preserves mirror reflection about $k_w$ and two zero-energy modes are located at the corners with mirror reflection about $z = -x$ and $y = 0$ planes in real space. Likewise, if we take $c_x = -c_z$, the system hosts $\mathcal{M}_{(z-x)yw}$-symmetry enforcing the position of two zero-modes located at the corners with mirror reflection about $z = x$ and $y = 0$ planes in real space, as shown in Fig. S8(b). We leave the relevant discussions for the interested readers.

### 3. 3D higher-order topological subsystems with varying $k_w$

Again, if we treat $k_w$ as a parameter $\varphi$ in $\mathcal{H}_2$ with $c_x = c_z = 0$, then we can repeat this story as we did in Sec. S-III A 3 where the hinge states becomes helical and the surface states become two-fold degenerate massive Dirac cones hosting the half value of spin Chern number. When we tune on $c_x = c_z$ being non-zero, 3D subsystems with fixing $\varphi$ between boundary nodal lines are the third-order TIs with two corner states.

## B. $\mathcal{H}_2^{k_z=0}$ in class AIII with spin conservation

### 1. Class AIII with spin conservation in 3D

By setting $k_z = 0$ in $\mathcal{H}_2$, we obtain a 3D novel TI phase $\mathcal{H}_2(k_x, k_y, k_w)$ for $b_{zw} = c_x = c_z = 0$ in class AIII. The system just two copies of $\mathcal{H}_1^{k_z=0}$, i.e., $\mathcal{H}_2^{k_z=0} = \sigma_3 \otimes \mathcal{H}_1^{k_z=0}$. The first-order boundary states become 2D double Dirac cones characterized by the 3D winding number $w_3$ defined in Sec. S-II C while the second-order Fermi-arc states are now consisting of double zero-energy modes at each hinge along $k_w$.

Discussions of this model akin to Sec. S-III A 2 can be simply repeated for the case when only $b_{xw}/b_{yw}$ is non-zero or $b_{xw} = \pm b_{yw}$, where the second-order hinge states consisting of helical states characterized by the spin MFCNs and spin winding numbers (similar to Eq.(S87)) instead.

### 2. 2D second-order topological subsystems with varying $k_w$

By treating $k_w$ as a periodic parameter $\varphi$ in $\mathcal{H}_2^{k_z=0}$ for $b_{zw} = c_x = c_z = 0$, we can also repeat this story as we did in Sec. S-III B 2. In the SOTI phase with zero-modes located at two corners, each corner hosts two modes.

## C. Nodal-line/sphere semimetal phases and field-theoretical descriptions

For a large value of $b_{xw}/b_{yw}/b_{zw}$ when $c_x = c_z = 0$, the bulk gap of $\mathcal{H}_2$ will close and become a nodal-line semimetal phase. Without loss of generality and for concreteness, we focus on the case when $M$ near 4 and $b_{xw}$ being small, one can expand $\mathcal{H}_2$ near the origin of BZ, then we have

$$\mathcal{H}_{eff} = k_x\tilde{\Gamma}_1 + k_y\tilde{\Gamma}_2 + k_z\tilde{\Gamma}_3 + k_w\tilde{\Gamma}_4 + m_0\tilde{\Gamma}_0 + b_{xw}\tilde{\Gamma}_0\tilde{\Gamma}_{14} \tag{S89}$$

with the energy spectrum

$$E = \pm\sqrt{k_y^2 + k_z^2 + (\sqrt{k_x^2 + k_w^2 + m_0^2} \pm b_{xw})^2} \tag{S90}$$

where $m_0 = m + \frac{1}{2}(k_y^2 + k_z^2)$ with $m = M - 4$. As we see that there is a four-fold degenerate nodal-line structure of band crossing points satisfying the equation:

$$k_x^2 + k_w^2 = \rho^2, \text{ with } \rho = \sqrt{b_{xw}^2 - m^2} \tag{S91}$$

at $k_y = k_z = 0$ plane when $|b_{xw}| > |m|$. Note that this nodal-line lives on $\mathcal{M}_{yz}$-invariant plane can be characterized by the spin first Chern number $C_{1s} = (C_1^+ - C_1^-)/2$ defined in the 2D subsystems inside and outside the nodal-line as we mentioned in Sec. S-III A 2. Thus this nodal structure can be treated as stacking of 2D spin Hall insulators in $k_x$-$k_w$ plane with a band crossing region forming a nodal-loop seperates the non-trivial and trivial phases inside and outside the nodal line respectively. This nodal line carry a $\mathbb{Z}_2$ nature protected by the $\tilde{\Gamma}_5$ symmetry with the effective continuum Hamiltonian is given by

$$H_{NL}^{4D,s} = k_z G_{31} - k_y G_{02} + (b_{xw} - \sqrt{k_x^2 + k_w^2 + m_0^2})G_{03}. \tag{S92}$$

The effective action can be simply generalized as spin version of Eq. (S62), taking the form,

$$S_{NL}^{4D,s} = \frac{\mathcal{B}_{ab}}{16\pi^3} \int (e^a \wedge e^b \wedge A^+ \wedge dA^+ - e^a \wedge e^b \wedge A^- \wedge dA^-), \tag{S93}$$

where $+(-)$ denotes spin up (down) component for each $4 \times 4$ Hamiltonian of $\mathcal{H}_2$. Similar discussion can be recovered as we did in Sec. S-III C. Likewise, by setting $k_z = 0$, we obtain the effective action of the nodal-line phase of $\mathcal{H}_2^{k_z=0}$, given by,

$$S_{NL}^{3D,s} = \frac{\mathcal{B}_{ab}}{8\pi^2} \int (e^a \wedge e^b \wedge dA^+ + e^a \wedge e^b \wedge dA^-). \tag{S94}$$

Note that the nodal line in Eq. (S92) with $k_z = 0$ indeed is the two same copies of the nodal-line with same charge.

In what follows we discuss the case when $b_{xw} = b_{yw} = b_{zw} = 0$. In this case, for a large value of $c_x/c_z$, we obtain a nodal sphere semimetal phase of $\mathcal{H}_2$. Without loss of generality and for concreteness, we focus on the case when $M$ near 4 and $c_x$ being small, one can expand $\mathcal{H}_2$ near the origin of BZ, then we have

$$\mathcal{H}_{eff} = k_x \tilde{\Gamma}_1 + k_y \tilde{\Gamma}_2 + k_z \tilde{\Gamma}_3 + k_w \tilde{\Gamma}_4 + m_0 \tilde{\Gamma}_0 + c_x \tilde{\Gamma}_{61} \tag{S95}$$

with the energy spectrum

$$E = \pm\sqrt{k_x^2 + (\sqrt{k_y^2 + k_z^2 + k_w^2 + m_0^2} \pm c_x)^2} \tag{S96}$$

where $m_0 = m + \frac{1}{2}k_x^2$ with $m = M - 4$. As we see that there is a four-fold degenerate nodal-sphere structure of band crossing points satisfying the equation:

$$k_y^2 + k_z^2 + k_w^2 = \rho^2, \text{ with } \rho = \sqrt{c_x^2 - m^2} \tag{S97}$$

at $k_x = 0$ plane when $|c_x| > |m|$. Note that this nodal-sphere lives on $\mathcal{M}_x$-invariant plane. Thus this nodal structure can be treated as stacking of 1D double SSH models in $(k_y, k_z, k_w)$ space with a band crossing region forming a nodal-loop seperates the non-trivial and trivial phases inside and outside the nodal-sphere respectively. This nodal line carry a $2\mathbb{Z}$ nature protected by chiral symmetry with the effective continuum Hamiltonian is given by

$$H_{NS}^{4D} = -k_x G_{20} + (c_x - \sqrt{k_y^2 + k_z^2 + k_w^2 + m_0^2})G_{30}. \tag{S98}$$

The codimenion of this nodal sphere is $d_c = 4 - 2 - 1 = 1$, thus we know that the system should be characterized by a $\mathbb{Z}$ number in class AIII for $d_c = 1$. We can define the topological number as

$$w = w_1^{in} - w_1^{out} = 2, \tag{S99}$$

where $w_1^{in} = 2$ ($w_1^{out} = 0$) denotes the 1D winding number for the middle effective four-band model inside (outside) the nodal sphere.

The effective action for this 4D nodal sphere can be derived in the same spirit by using the $LU(1)$ theory, taking the form,

$$S_{DNS} = \alpha \mathcal{B}_{abc} \int e^a \wedge e^b \wedge e^c \wedge dA, \tag{S100}$$

where $\alpha$ needs to be further determined below, and $e^a = e_\mu^a dx^\mu$ being the translation gauge field.

One can start by considering the $d = (4+1) + 2 + 1 = 8$ dimensional axion theory[11] with $\Theta = \pi$,

$$S = 2 \times \frac{\pi}{4!(2\pi)^4} \int d\mathcal{A} \wedge d\mathcal{A} \wedge d\mathcal{A} \wedge d\mathcal{A}, \tag{S101}$$

where the factor 2 comes from the double version of the 4D nodal sphere, i.e., $w = 2$. Inserting $\mathcal{A} = A + yK_y + zK_z + wK_w$ into the above action, we have

$$S = \frac{1}{4!(2\pi)^4} \int d(A+yK_y+zK_z+wK_w) \wedge d(A+yK_y+zK_z+wK_w) \wedge d(A+yK_y+zK_z+wK_w) \wedge d(A+yK_y+zK_z+wK_w). \tag{S102}$$

Here $y, z, w$ denote the translation gauge fields and $K_{y,z,w}(\theta,\varphi)$ are the coordinates of the points on the nodal sphere ($\theta \in [0,\pi], \varphi \in [0,2\pi]$) with the radius $\rho$ when only $c_x$ is non-zero. This gives rise to a mixed anomaly term, except in one extra dimension

$$S = 2 \times 4 \times \frac{\pi}{4!(2\pi)^4} \int [d(yK_y+zK_z+wK_w) \wedge d(yK_y+zK_z+wK_w) \wedge d(yK_y+zK_z+wK_w) \wedge dA], \tag{S103}$$

corresponding to a boundary theory (note that $d^2 = 0$)

$$S = \frac{\pi}{3(2\pi)^4} \int (yK_y+zK_z+wK_w) \wedge d(yK_y+zK_z+wK_w) \wedge d(yK_y+zK_z+wK_w) \wedge dA$$
$$= \frac{V_F}{8\pi^3} \int y \wedge z \wedge w \wedge dA, \tag{S104}$$

where $V_F = \frac{1}{3}\int d\theta d\varphi J_{yzw}(\theta,\varphi)$ is the volume in the momentum space enclosed by the nodal sphere with

$$J_{yzw}(\theta,\varphi) = \begin{vmatrix} K_y & K_z & K_w \\ \partial_\theta K_y & \partial_\theta K_z & \partial_\theta K_w \\ \partial_\varphi K_y & \partial_\varphi K_z & \partial_\varphi K_w \end{vmatrix}. \tag{S105}$$

For instance, if we assume $K_y = \rho \sin\theta \cos\varphi$, $K_z = \rho \sin\theta \sin\varphi$, and $K_w = \rho \cos\theta$, then we have $V_F = \frac{4}{3}\pi\rho^3$ being the volume of a standard 3D ball enclosed by the two-sphere.

By replacing $y \to e^y$, $z \to e^z$, $w \to e^w$, and $V_F \to \mathcal{B}_{yzw}$, we obtain the action as presented in Eq. (S106). The time-component term comes from $\mathcal{A} = A + yK_y + zK_z + wK_w + tK_t$, where the nodal sphere now tilts along $k_y$ direction in energy $E$. Then we obtain $\mathcal{B}_{tzw} = \frac{1}{3}\int d\theta d\varphi J_{tzw}(\theta,\varphi)$, which is nothing but the projection volume enclosed by the nodal sphere in $(E, k_z, k_w)$ space. For $e_y^y = 1$, $e_z^z = 1$, $e_w^w = 1$, and $e_t^t = 1$, we have

$$S_{DNS} = \frac{\mathcal{B}_{yzw}}{8\pi^3} \int d^5 x \epsilon^{yzw\mu\nu} \partial_\mu A_\nu + \frac{\mathcal{B}_{tzw}}{8\pi^3} \int d^5 x \epsilon^{tzw\mu\nu} \partial_\mu A_\nu, \tag{S106}$$

where $x^\mu = (t, x, y, z, w)$. The first term can be undersood as the summation of (1+1)D axion theory in 3D momentum space $(k_y, k_z, k_w)$ for the effective 1D topological insulator $\mathcal{H}_2(k_x)$ hosting a theta term equal $2\pi$. Alternately, for fixed $(\theta, \varphi)$ on the sphere, there is a pair of 2D double Dirac points tilted in energy along $k_y$ axis inducing (2+1)D parity anomaly[19], thus the first and second terms are the summation of this effects when $(\theta, \varphi)$ run over the whole sphere. By setting $k_z = 0$ in $\mathcal{H}_2$ in the nodal sphere case, the nodal structure becomes a 3D double-nodal-line, taking the action as

$$S_{DNL} = \frac{\mathcal{B}_{yw}}{4\pi^2} \int d^4 x \epsilon^{yw\mu\nu} \partial_\mu A_\nu + \frac{\mathcal{B}_{tw}}{4\pi^2} \int d^4 x \epsilon^{tw\mu\nu} \partial_\mu A_\nu, \tag{S107}$$

with $\mathcal{B}_{yw}$ denotes the projection area in $k_y$-$k_w$ plane enclosed the nodal line while $\mathcal{B}_{tw}$ denotes the projection area in $E$-$k_w$ plane while tilting this nodal line along $k_y$ axis.

## S-V.  $\mathcal{H}_3$ IN CLASS A$^{C_4\mathcal{T}}$ AND $\mathcal{H}_3^{k_z=0}$ IN CLASS AIII$^{C_4\mathcal{T}}$

### A.  $\mathcal{H}_3$ in Class A$^{C_4\mathcal{T}}$

#### 1.  Shifted boundary Weyl cones

The first-order boundary states can be obtained akin to the method in Sec. S-III A 1. One can project $\mathcal{H}_0$ by $P_\alpha^\pm$ and obtain boundary Weyl cones at the origin of the first-order boundary BZ. And then the perturbation $\Delta_4 = \delta(\cos k_x - \cos k_y)\Gamma_4$ in the continuum limit near the origin when $2 < M < 4$ presented as

$$\Delta_4^x = \delta\frac{1}{2}\partial_x^2\Gamma_4,$$
$$\Delta_4^y = -\delta\frac{1}{2}\partial_y^2\Gamma_4,$$

(S108)

are projected into the boundaries as

$$P_x^\pm\Delta_4^x P_x^\pm = \frac{\delta}{2}\int_0^{+\infty}dx\varphi^*(x)\partial_x^2\varphi(x)P_x^\pm\Gamma_4 P_x^\pm = -v_w P_x^\pm\Gamma_4 P_x^\pm,$$
$$P_y^\pm\Delta_4^y P_y^\pm = -\frac{\delta}{2}\int_0^{+\infty}dy\varphi^*(y)\partial_y^2\varphi(y)P_y^\pm\Gamma_4 P_y^\pm = v_w P_y^\pm\Gamma_4 P_y^\pm,$$

(S109)

where $\varphi(x) \sim e^{-\lambda x}$ and $\varphi(y) \sim e^{-\lambda y}$ with $\lambda$ being positive are the spatial part of the eigenstates of differential equations

$$\left[\partial_x + (m - \frac{1}{2}\partial_x^2)i\Gamma_1\Gamma_0\right]\psi(x) = 0,$$
$$\left[\partial_y + (m - \frac{1}{2}\partial_y^2)i\Gamma_2\Gamma_0\right]\psi(y) = 0,$$

(S110)

with $m = M - 4$ when considering OBC along $x$ and $y$ directions respectively. Above we take $\psi(x) = \chi_\eta^x\varphi(x)$ and $\psi(y) = \chi_\eta^y\varphi(y)$ with $\chi_\eta^x$ and $\chi_\eta^y$ being the eigenvectors of $i\Gamma_1\Gamma_0$ and $i\Gamma_2\Gamma_0$ corresponding to $\eta = \pm 1$ respectively. For these detailed derivations, we refer to Ref. [20]. Thus, the low-energy noundary Hamiltonian of $\mathcal{H}_3$ near the origin for $2 < M < 4$ is given by

$$\mathcal{H}_{x,\pm} = \mp k_y\sigma_2 \mp k_z\sigma_3 \pm (k_w - v_w)\sigma_1,$$
$$\mathcal{H}_{y,\pm} = \mp k_x\sigma_1 \mp k_z\sigma_3 \pm (k_w + v_w)\sigma_2,$$

(S111)

respectively. At the boundary $x = \pm L/2$, there is a Weyl cone located at $\mathbf{k}_{x\pm} = (k_y, k_z, k_w) = (0, 0, v_w)$ with chirality $\chi_\pm = \pm 1$; At $y = \pm L/2$, there is a Weyl cone located at $\mathbf{k}_{y\pm} = (k_x, k_z, k_w) = (0, 0, -v_w)$ with chirality $\chi_\pm = \pm 1$, as illustrated in numeric results in Fig. S10. Note that $\Delta_4$ does not affect the boundary Weyl cones in the case of OBCs at $z$ and $w$ directions.

#### 2.  Topological invariants

For $\mathcal{H}_3$ preserves $C_4^{zw}\mathcal{T}$-symmetry, we can define an extra topological invariant at two $C_4^{zw}\mathcal{T}$-invariant 2D momenta $\Lambda_{1,2} = (k_x, k_y) = (0, 0), (\pi, \pi)$. At these two points, the subsystem $\mathcal{H}_3^{\Lambda_a}(k_y, k_z)$ with $a = 1, 2$ hosts $\mathcal{T}$ and $\Gamma_5$-symmetries, i.e., $[\Gamma_5, \mathcal{H}_3^{\Lambda_a}] = 0$, then we define the topological numbers at this two momenta as the "rotational first Chern number" $C_{1r}(\Lambda_a) = (C_+ - C_-)/2$ where $C_\pm$ denotes the first Chern number of the $2 \times 2$ block Hamiltonian in the eigenspace of $\mathcal{E} = \pm 1$ of $\hat{\Gamma}_5 = G_{03}$, which may imply the generation of the second-order Fermi-arcs located at four-hinges in this phase when $\Delta_4$ is introduced.

In order to gain a deeper understanding of the hinge modes, we will focus on the combined mirror symmetries of this model, specifically the $\mathcal{M}_{xy}$, $\mathcal{M}_{yz}$, and $\mathcal{M}_{xz}$ symmetries. These symmetries use the same operators as discussed in the previous section, with $\hat{\mathcal{M}}_{ij} = \Gamma_{ij}$. Together with $\mathcal{M}_{(x-y)w}$ and $\mathcal{M}_{(x+y)w}$ defined as $\mathcal{M}_{(x-y)w} = \mathcal{M}_x C_4^{zw}\mathcal{M}_w$, and $\mathcal{M}_{(x+y)w} = \mathcal{M}_y C_4^{zw}\mathcal{M}_w$ with the operators $\hat{\mathcal{M}}_{(x-y)w} = \Gamma_1\Gamma_4 e^{-\frac{\pi}{4}\Gamma_1\Gamma_2}$, and $\hat{\mathcal{M}}_{(x+y)w} = \Gamma_2\Gamma_4 e^{-\frac{\pi}{4}\Gamma_1\Gamma_2}$. Please remind the definition of combined mirror symmetry in Sec. S-I.

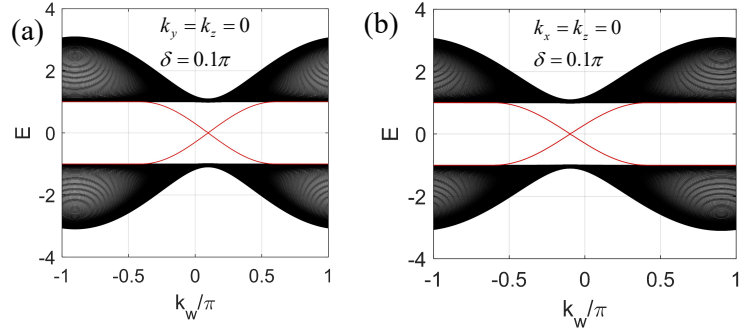

FIG. S10. Energy spectrum when considering OBC along (a) $x$ direction for $k_y = k_z = 0$; (b) $y$ direction for $k_x = k_z = 0$. We take $\delta = 0.1\pi$ and $L = 80$. As we can see that the Weyl cones are shifted along $k_w$ direction at $k_w = v_w$ ($k_w = -v_w$) at $x = \pm L/2$ ($y = \pm L/2$) with $v_w \approx \delta$ when $\delta$ being numeric small.

Without loss of generality and for simplicity, we consider the case when $M = 3$. Evaluation shows that $C_{1r}(\Lambda_1) = 1$, and $C_{1r}(\Lambda_2) = 0$ which implies the existence of the second-order Fermi-arcs at $k_z = 0$ plane. For $\mathcal{M}_{ij}$-symmetry , we have the MFCNs $C_{1m}^{xy}(\Lambda_1^{xy}) = 1$, $C_{1m}^{yz}(\Lambda_1^{yz}) = 1$, $C_{1m}^{xz}(\Lambda_1^{xz}) = 1$ and the MFCNs are zero at other high-symmetry points when $\delta$ being numeric small. It means that one of the hinge modes will be transformed to a $\mathcal{M}_{xy}$-symmetric hinge about $x = 0$ and $y = 0$ planes with the same chirality; those hinge modes at $y(x)$-$k_z$ plane will be transformed to the corresponding $\mathcal{M}_y(\mathcal{M}_x)$-symmetric hinges about $y = 0$ ($x = 0$) planes and changes its chirality due to the action of $\mathcal{M}_z$, as shown in Fig. S11(a-c). In addition, the non-zero MFCNs protected by $\mathcal{M}_{(x-y)w}$ and $\mathcal{M}_{(x+y)w}$ symmetries indicate the hinge modes located at diagonal and anti-diagonal hinges at $x$-$y$ planes satisfying the $\mathcal{M}_w$-symmetry about $k_w = 0$, respectively. All these calculations of MFCNs can be recovered in the same spirit as we did in Sec. S-III A 2.

Note that an extra perturbation $\delta_1 = \lambda\Gamma_4$ breaks $C_4^{zw}\mathcal{T}$, and $\mathcal{M}_{(x\pm y)w}$ symmetries, the system still keeps the second-order Fermi-arc states but just shifts the position of Fermi-arc along $k_w$-axis, which is now protected by $\mathcal{M}_{xy}$, $\mathcal{M}_{yz}$, $\mathcal{M}_{xz}$-symmetries. If we further introduce the previous perturbation $\Delta_1$ with only $b_{xw} \neq 0$, it breaks $\mathcal{M}_{xy}$, and $\mathcal{M}_{xz}$ which still hosts shifted second-order Fermi-arcs at the hinges protected by $\mathcal{M}_{yz}$-symmetry. Again, we emphasize that $C_4^{zw}\mathcal{T}$ and $\mathcal{M}_{ij}$-symmetries are used to construct this novel TI phases, these hinge states are robust against the corresponding symmetry breaking perturbations, as long as the bulk gap is not closed.

Similarly, this 4D novel phase can be regarded as the stacking of a series of 3D chiral SOTIs by treating $k_w$ as a parameter $\varphi$. The 3D subsystems $\mathcal{H}_2^\varphi(k_x, k_y, k_z)$ restore an extra charge-conjugate symmetry with $\mathcal{C} = G_{22}\mathcal{K}$ satisfying $\mathcal{C}^2 = +1$, and thus falls into class D. Fortunately, it still preserves the corresponding $\mathcal{M}_{ij}$-symmetry as in 4D. One can discuss its second-order chiral hinge states from the first-order boundaries as in Sec. S-III A 3.

## B. $\mathcal{H}_3^{k_z=0}$ in class AIII$^{C_4\mathcal{T}}$

In the 3D case when $k_z = 0$, the system $\mathcal{H}_1(k_x, k_y, k_w)$ with the mass $m = M - 1$ restores chiral symmetry with $\mathcal{S} = \Gamma_3$ and thus falls into class AIII. The model preserves $C_4^w\mathcal{T}$ with the same operator as in 4D. Meanwhile, the system hosts two extra mirror symmetries with $\hat{\mathcal{M}}_x = \Gamma_1\Gamma_3$, and $\hat{\mathcal{M}}_y = \Gamma_2\Gamma_3$. One can define the MFCNs $C_{1m}^{k_x}/C_{1m}^{k_y}$ at the mirror invariant planes $k_x/k_y = 0, \pi$, i.e. $C_{1m} = (C_{+i} - C_{-i})/2$, to characterize the second-order Fermi arcs. For instance, when $M = 3$ with $m_a = m - \cos k_x$ and $\delta = 0.1\pi$, we obtain $C_{1m}^0 = -1$, and $C_{1m}^\pi = 0$ for $\mathcal{M}_x$. In Fig. S11(d), we present the numeric results of first Chern number of each mirror invariant block Hamiltonian with the eigenvalue $\epsilon = \pm i$ of $\hat{\mathcal{M}}_x$. Similarly, we have $C_{1m}^0 = 1$, and $C_{1m}^\pi = 0$ for $\mathcal{M}_y$ when $M = 3$ and $\delta = 0.1\pi$. The picture of the boundary states of this 3D novel TI is similar to the 4D case as we mentioned above. The difference is that this phase is now stacking by a series of 2D SOTIs along $k_w$ axis where the zero-energy corner states form a second-order Fermi-arc at four hinges.

## S-VI. $\mathcal{H}_4$ IN CLASS AIII$^{C_4\mathcal{T}}$ AND $\mathcal{H}_4^{k_z=0}$ IN CLASS AIII$^{C_4\mathcal{T}}$ WITH SPIN CONSERVATION

Note that $\mathcal{H}_4 = \sigma_3 \otimes \mathcal{H}_3$ is just two copies of $\pm\mathcal{H}_3$ with the opposite second Chern numbers in the bulk. The system preserves an extra $\bar{\Gamma}_5$-symmetry, i.e., with spin conservation. The first-order boundary states become Dirac

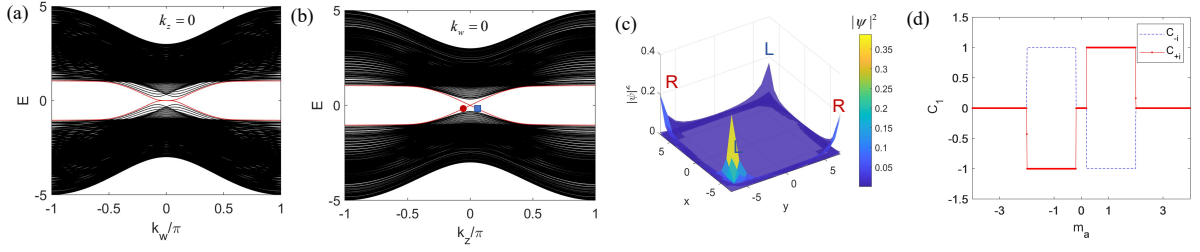

FIG. S11. Energy spectrum when considering OBC along $x$ and $y$ directions of $\mathcal{H}_3$. (a)$E(k_w)$ when $k_z = 0$; (b) $E(k_z)$ when $k_w = 0$. The location of four chiral hinge modes:(c) when $k_w = 0$, and $k_z = \pm 0.1$; "L" and "R" denotes left and right chirality of the hinge modes respectively. (d)First Chern number of each mirror-invariant block Hamiltonian $\mathcal{H}_\pm$ at $k_x = 0, \pi$ in the eigenspace $\epsilon = \pm i$ of $\tilde{\mathcal{M}}_x$ where $m_a = m - \cos k_x$ with $a = 1, 2$ corresponding to $k_x = 0, \pi$, respectively. In the above numeric, we set the lattice length $L = 15$, $M = 3$, and $\delta = 0.1\pi$.

cones while the four second-order hinge states become helical. All the related topological numbers can be simply generalized into the spin version as half difference of the topological numbers defined in two block systems as usual.

For $\mathcal{H}_4^{k_z=0}$ in Class AIII$^{C_4\mathcal{T}}$, it consists of two 3D subsystems $\pm\mathcal{H}_3^{k_z}$ with the same 3D winding numbers in the bulk. The system preserves an extra $\tilde{\Gamma}_5$-symmetry, i.e., with spin conservation. The first-order boundary states become double Dirac cones while the second-order Fermi-arcs now consists of two-fold degenerate zero-modes along $k_w$ axis at each hinge. All the related topological numbers can be simply generalized into the spin version as half difference of the topological numbers defined in two block systems as usual.

# S-VII. NOVEL TOPOLOGICAL SEMIMETALS WITH HYBRID-ORDER FERMI-ARCS/SURFACES

## A. $\mathcal{H}_1/\mathcal{H}_1^{k_z=0}$ in 5D/4D

We now stack a 4D novel TI $\mathcal{H}_1$ along an additional $u$-dimension by replacing $M$ with $M' = M - \cos k_u$. Without loss of generality and for simplicity, we only focus on the case when $M = 4$ in this section. Before introducing $\Delta_1$, the system now hosts a pair of 5D Yang monopoles [21–23] located at $\boldsymbol{K}_\pm = (k_x, k_y, k_z, k_w, k_u) = (0, 0, 0, 0, \pm\frac{\pi}{2})$ with opposite charges labeled by $C_2 = \pm 1$ defined on the four-sphere enclosed the monopole. Near $\boldsymbol{K}_\pm$, we have the low-energy Hamiltonian, given by,

$$\mathcal{H}_{YM}^\pm = q_x\Gamma_1 + q_y\Gamma_2 + q_z\Gamma_3 + q_w\Gamma_4 + q_u\Gamma_0,$$ (S112)

where $\boldsymbol{q} = \boldsymbol{k} - \boldsymbol{K}_\pm$. Once we introduce the perturbation term $b_{xw}\Gamma_0\Gamma_{14}$ term ($b_{xw} > 0$) into $\mathcal{H}_1$, each monopole expands into a nodal sphere with the effective Hamiltonian

$$\mathcal{H}_{YS}^\pm = \mathcal{H}_{YM}^\pm + b_{xw}\Gamma_0\Gamma_{14},$$ (S113)

with the energy spectrum

$$E = \pm\sqrt{q_y^2 + q_z^2 + (\sqrt{q_x^2 + q_w^2 + q_u^2} \pm b_{xw})^2}.$$ (S114)

Thus we find the band crossing points forming a nodal sphere with two-fold degeneracy satisfying the equation

$$q_x^2 + q_w^2 + q_u^2 = b_{xw}^2,$$ (S115)

at $q_y = q_z = 0$ plane. One can simply obtain the effective model for middle two-bands of these nodal spheres as we did before [6], with the Hamitonian

$$\mathcal{H}_{NS,\pm}^{5D} = q_z\sigma_1 - q_y\sigma_2 + \left(b_{xw} - \sqrt{q_x^2 + q_w^2 + \left[\frac{1}{2}(q_y^2 + q_z^2) + q_u\right]^2}\right)\sigma_3.$$ (S116)

Note that this nodal sphere actually host a double charge $Q_D = (C_2, \Delta_1)$. In this case, $C_2$ is still unchanged defined on the four-sphere enclosed this nodal-surface while $\Delta_1 = C_1^{in} - C_1^{out}$ is the difference between the first

Chern number defined in the occupied bands of the 2D subsystems $\mathcal{H}_{\boldsymbol{q}_\perp}(\boldsymbol{q}_\parallel)$ inside and outside the nodal sphere, i.e., $q_\perp < b_{xw}$ and $q_\perp > b_{xw}$ respectively. Here $q_\perp = \sqrt{q_x^2 + q_w^2 + q_u^2}$ and $q_\parallel = \sqrt{q_y^2 + q_z^2}$. Calculation shows that $\mathcal{H}_{YS}^\pm$ hosts a double charge $(\pm 1, 1)$. We refer to Sec. S-III A 2 for calculation details. In fact, 2D subsystems in the lattice model are the non-trivial Chern insulators for $q_\perp < b_{xw}$ while those are trivial for $q_\perp > b_{xw}$. This semimetal phase harbors both first-order Fermi-arcs and second-order Fermi-surfaces due to the stacking effect of 4D TIs with hybrid-order boundary states between two nodal spheres.

By setting $k_z = 0$, we obtain the 4D semimetal of $\mathcal{H}_1^{k_z=0}$ hosting a pair of 4D nodal spheres with the Hamiltonian in Eq. (S113) with $k_z = 0$ labelled as $\mathcal{H}_{DS}^\pm$. These nodal spheres are now at $k_y = 0$ plane hosting a double charge $Q_D = (w_3, \Delta w_1)$. In this case, $w_3$ is the 3D winding number defined on the three-sphere enclosed the 4D nodal-surface with the same result to the monopole case with $b_{xw}$ vanishes. $\Delta w_1 = w_1^{in} - w_1^{out}$ is the difference between the 1D winding number of the 1D subsystems $\mathcal{H}_{\boldsymbol{q}_\perp}(q_y)$ defined inside and outside the nodal sphere, i.e., $q_\perp < b_{xw}$ and $q_\perp > b_{xw}$ respectively. Calculation shows that $\mathcal{H}_{DS}^\pm$ hosts a double charge $(\pm 1, 1)$. We refer to Sec. S-III B 1 for calculation details. In fact, 1D lattice subsystems are the non-trivial SSH models for $q_\perp < b_{xw}$ while those are trivial for $q_\perp > b_{xw}$. This semimetal phase harbors both first-order and second-order Fermi-surfaces due to the stacking effect of 3D TIs with hybrid-order boundary states between two nodal spheres.

### B. $\mathcal{H}_2/\mathcal{H}_2^{k_z=0}$ in 5D/4D

For $c_x = c_z = 0$, $\mathcal{H}_2$ is just two copies of $\pm\mathcal{H}_1$. A similar disucssion can be recovered for the case when $M = 4$ and only $b_{xw}$ survives. The nodal-sphere becomes a spin version of $\mathcal{H}_{YS}$, we can use the spin second-Chern and spin first-Chern numbers $C_{ns} = (C_n^+ - C_n^-)/2$ with $n = 1, 2$ to describe the double charge of these spin Yang nodal-spheres accordingly. By setting $k_z = 0$, we use the spin winding numbers $w_{ns} = (w_n^+ - w_n^-)/2$ with $n = 1, 3$ to characterize these 4D nodal spheres.

### C. $\mathcal{H}_3/\mathcal{H}_4$ and $\mathcal{H}_3^{k_z=0}/\mathcal{H}_4^{k_z=0}$ in 5D and 4D

By stacking $\mathcal{H}_3$ along $u$ dimension as we did above, we obtain a second-order 5D Yang semimetal consisting of a pair of Yang monopoles when $M = 4$ at $\boldsymbol{K}_\pm$. To distinguish them from the conventional Yang monopole, we expand the effective low-energy Hamiltonian of these monopole as

$$\mathcal{H}_{SOYM}^\pm = q_x\Gamma_1 + q_y\Gamma_2 + q_z\Gamma_3 + (q_w + \frac{\delta}{2}(q_y^2 - q_x^2))\Gamma_4 \pm q_u\Gamma_0, \tag{S117}$$

which still hosts the same charge $C_2 = \pm 1$ as in the case when $\delta = 0$. Th system now hosts both first-order Fermi-arcs and second-order Fermi-surfaces due to the stacking effect of 4D TI $\mathcal{H}_3$ along $k_u$ axis. The semimetal phase of $\mathcal{H}_4$ is just two copies of $\pm\mathcal{H}_3$ while these monopoles are just characterized by $C_{2s}$. Similar discussions can be repeated by setting $k_z = 0$ for the semimetal phase of $\mathcal{H}_3^{k_z=0}/\mathcal{H}_4^{k_z=0}$. We no longer repeat this story here.

## S-VIII. REALIZATION PROPOSAL USING ULTRACOLD ATOMS

Here we present the proposal for realizing the four-band Hamiltonian of the main text. The eight-band model in the main text can refer the previous work [6]. We start with the degenerate Fermi gases trapped in the optical lattice. In order to generate such a model, we use four hyperfine states $|g_{1,2}\rangle$ and $|e_{1,2}\rangle$ of the atoms as the pseudo-spins. The four hyperfine states are coupled by optical fields between the $|g\rangle$ and $|e\rangle$ manifolds. The setup that engineer the Hamiltonian is illustrated in Fig.S12.

The Hamiltonian of the system is expressed in two parts,

$$\mathcal{H}_0 = \hat{H}_0 + \hat{H}_{so}. \tag{S118}$$

The first part describes the local energy of the atoms,

$$\hat{H}_0 = \int d\mathbf{r} \sum_{\lambda,\lambda'} \psi_\lambda^\dagger(\mathbf{r}) \left\{ \left[ -\frac{\nabla^2}{2m} + \hat{V}_{\text{trap}}(\mathbf{r}) \right] \delta_{\lambda\lambda'} + \hat{V}[\sigma_z \otimes \sigma_0]_{\lambda,\lambda'} \right\} \psi_{\lambda'}(\mathbf{r}). \tag{S119}$$

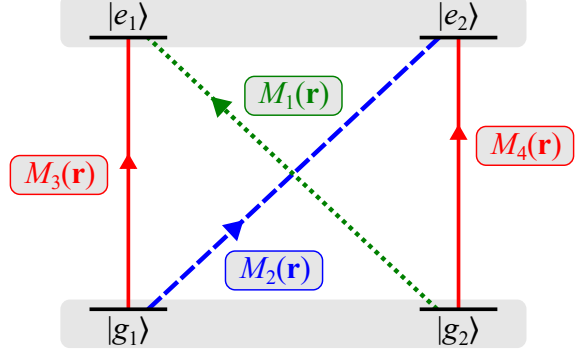

FIG. S12. Illustration of the experimental setups. We use $^{87}$Sr as the example [24] and choose its state manifolds $^1S_0$ and $^3P_0$ as $|g\rangle$ and $|e\rangle$. The red-solid, blue-dashed, and green-dotted lines show the coupling with field modes of $M_{3,4}(\mathbf{r})$, $M_2(\mathbf{r})$, and $M_1(\mathbf{r})$. They are respectively generated by the $\pi$, $\sigma^+$, and $\sigma^-$-polarized fields.

Here $\psi_\lambda$ stands for the atomic operator with $\lambda = e_1, e_2, g_1, g_2$ denoting the pseudo-spins. $\hat{V}_{\text{trap}} = V_L \sin^2(k_L x) + V_L \sin^2(k_L y) + V_L \sin^2(k_L z)$ is the lattice trap potential, and $\hat{V}$ describes the energy difference between the $|g\rangle$ and $|e\rangle$ manifolds. The second part describes the coupling between the pseudo-spin $|g\rangle$ and $|e\rangle$ manifolds,

$$\hat{H}_{\text{so}} = \int d\mathbf{r} \Big[ M_1(\mathbf{r})\psi_{e_1}^\dagger(\mathbf{r})\psi_{g_2}(\mathbf{r}) + M_2(\mathbf{r})\psi_{e_2}^\dagger(\mathbf{r})\psi_{g_1}(\mathbf{r}) + M_3(\mathbf{r})\psi_{e_1}^\dagger(\mathbf{r})\psi_{g_1}(\mathbf{r}) + M_4(\mathbf{r})\psi_{e_2}^\dagger(\mathbf{r})\psi_{g_2}(\mathbf{r}) \Big] + H.c. \quad \text{(S120)}$$

where the standing-wave modes are given as follows,

$$\begin{cases} M_1(\mathbf{r}) = iM_1 \sin(k_x d)\cos(k_y d)\cos(k_z d) + M_1' \cos(k_x d)\sin(k_y d)\cos(k_z d) \\ M_2(\mathbf{r}) = iM_2 \sin(k_x d)\cos(k_y d)\cos(k_z d) - M_2' \cos(k_x d)\sin(k_y d)\cos(k_z d) \\ M_3(\mathbf{r}) = iM_3 \cos(k_x d)\cos(k_y d)\sin(k_z d) - iM_3' \cos(k_x d)\cos(k_y d)\cos(k_z d) \\ M_4(\mathbf{r}) = -iM_4 \cos(k_x d)\cos(k_y d)\sin(k_z d) - iM_4' \cos(k_x d)\cos(k_y d)\cos(k_z d) \end{cases}. \quad \text{(S121)}$$

We expand $\psi_\lambda(\mathbf{r})$ in terms of the Wannier wave function $W(\mathbf{r})$,

$$\psi_\lambda(\mathbf{r}) = \sum_j W(\mathbf{r} - \mathbf{r}_j)\psi_{j\lambda}, \quad \text{(S122)}$$

where $j$ denotes the site index and $\mathbf{r}_j$ the coordinate of the site center. We use the tight-binding approximation, then the Hamiltonian (S118) is given as

$$\hat{H}_0 = \sum_j \Big[ -J\psi_j^\dagger(\sigma_0 \otimes \sigma_0)\psi_{j+1} + H.c. \Big] + \hat{V}\psi_j^\dagger(\sigma_z \otimes \sigma_0)\psi_j, \quad \text{(S123)}$$

$$\begin{aligned} \hat{H}_{\text{so}} = \sum_j (-1)^{j_x+j_y+j_z} \Big\{ &(iA_x)(\psi_{j,e_1}^\dagger\psi_{j+\mathbf{e}_x,g_2} + \psi_{j,e_2}^\dagger\psi_{j+\mathbf{e}_x,g_1} - \psi_{j,e_1}^\dagger\psi_{j-\mathbf{e}_x,g_2} - \psi_{j,e_2}^\dagger\psi_{j-\mathbf{e}_x,g_1}) \\ &+ A_y(\psi_{j,e_1}^\dagger\psi_{j+\mathbf{e}_y,g_2} - \psi_{j,e_2}^\dagger\psi_{j+\mathbf{e}_y,g_1} - \psi_{j,e_1}^\dagger\psi_{j-\mathbf{e}_y,g_2} + \psi_{j,e_2}^\dagger\psi_{j-\mathbf{e}_y,g_1}) \\ &+ \sum_{\nu=1,2} \Big[ (iA_z)[\sigma_z]_{\nu\nu'}(\psi_{j,e_\nu}^\dagger\psi_{j+\mathbf{e}_z,g_{\nu'}} - \psi_{j,e_\nu}^\dagger\psi_{j-\mathbf{e}_z,g_{\nu'}}) + (-iA_w)\psi_{j,e_\nu}^\dagger\psi_{j,g_\nu} \Big] \Big\} + H.c. \end{aligned} \quad \text{(S124)}$$

where we have defined $\psi = (\psi_{e_1}, \psi_{e_2}, \psi_{g_1}, \psi_{g_2})^T$, and tuned the optical-field strength as

$$\begin{cases} A_x = M_{1,2} \int \sin(k_x d)\cos(k_y d)\cos(k_z d)W^*(\mathbf{r})W(\mathbf{r}+\mathbf{e}_x)d\mathbf{r} \\ A_y = M_{1,2}' \int \cos(k_x d)\sin(k_y d)\cos(k_z d)W^*(\mathbf{r})W(\mathbf{r}+\mathbf{e}_y)d\mathbf{r} \\ A_z = M_{3,4} \int \cos(k_x d)\cos(k_y d)\sin(k_z d)W^*(\mathbf{r})W(\mathbf{r}+\mathbf{e}_z)d\mathbf{r} \\ A_w = M_{3,4}' \int \cos(k_x d)\cos(k_y d)\cos(k_z d)|W(\mathbf{r})|^2 d\mathbf{r} \end{cases} \quad \text{(S125)}$$

We note that in obtaining the $A_{x,y,z}$ of Eq.(S124), the on-site terms vanish because of the odd parity of $M_{1,2,3}(\mathbf{r})$. Thereby the nearest-neighbor terms dominate. Since there exists a staggered sign factor in Eq.(S124), we make the following operator representation that preserves the anti-commutation relation of $\psi_{j\lambda}$,

$$\psi_{j,g_\nu} \to (-1)^{j_x+j_y+j_z}\psi_{j,g_\nu} \tag{S126}$$

Under the transformation (S126), we obtain the Hamiltonian as

$$\hat{H}_0 = \sum_j \Big[ -J\psi_j^\dagger(\sigma_z \otimes \sigma_0)\psi_{j+1} + H.c. \Big] + \hat{V}\psi_j^\dagger(\sigma_z \otimes \sigma_0)\psi_j \tag{S127}$$

$$\hat{H}_{\mathrm{so}} = \sum_j (iA_x)[\psi_{j,e_1}^\dagger \psi_{j+\mathbf{e}_x,g_2} + \psi_{j,e_2}^\dagger \psi_{j+\mathbf{e}_x,g_1} - \psi_{j,e_1}^\dagger \psi_{j-\mathbf{e}_x,g_2} - \psi_{j,e_2}^\dagger \psi_{j-\mathbf{e}_x,g_1}]$$

$$+ A_y[\psi_{j,e_1}^\dagger \psi_{j+\mathbf{e}_y,g_2} - \psi_{j,e_2}^\dagger \psi_{j+\mathbf{e}_y,g_1} - \psi_{j,e_1}^\dagger \psi_{j-\mathbf{e}_y,g_2} + \psi_{j,e_2}^\dagger \psi_{j-\mathbf{e}_y,g_1}]$$

$$+ \sum_{\nu=1,2} \Big[ (iA_z)[\sigma_z]_{\nu\nu'}(\psi_{j,e_\nu}^\dagger \psi_{j+\mathbf{e}_z,g_{\nu'}} - \psi_{j,e_\nu}^\dagger \psi_{j-\mathbf{e}_z,g_{\nu'}}) + (-iA_w)\psi_{j,e_\nu}^\dagger \psi_{j,g_\nu} \Big] + H.c. \tag{S128}$$

In the momentum space, it gives

$$\mathcal{H}_0(\mathbf{k}) = \hat{H}_0(\mathbf{k}) + \hat{H}_{\mathrm{so}}(\mathbf{k}) \tag{S129}$$

$$\hat{H}_0(\mathbf{k}) = d_0(\mathbf{k})\sigma_z \otimes \sigma_0 \,, \quad \begin{cases} d_0(\mathbf{k}) = \hat{V} - 2J\sum_{i=x,y,z}\cos(k_id) \equiv M - 2J\sum_{i=x,y,z,w}\cos(k_id) \\ M = \hat{V} + 2J\cos(k_wd) \end{cases} \tag{S130}$$

$$\hat{H}_{\mathrm{so}}(\mathbf{k}) = \sum_{i=x,y,z} d_i(\mathbf{k})\sigma_x \otimes \sigma_i + d_4(\mathbf{k})\sigma_y \otimes \sigma_0 \,, \quad \begin{cases} d_i(\mathbf{k}) = 2A_i\sin(k_id) \\ d_4(\mathbf{k}) = A_w \equiv A_w'\sin(k_wd) \end{cases} \tag{S131}$$

We note that the parameter $k_w$ is recognized as the momentum of the synthetic dimension. After making $\sigma_x \leftrightarrow \sigma_z$ in the first $\sigma$-matrix space, we can find Hamiltonian (S129) is identical to $\mathcal{H}_0$ of Hamiltonian (1) in the main text.

The perturbation term $\Delta_1$ can be introduced by optical fields that generate the intra-manifold coupling of $|g\rangle$ or $|e\rangle$,

$$\hat{H}_1 = \int d\mathbf{r} \sum_{\lambda=g,e} \Omega \psi_{\lambda_1}^\dagger(\mathbf{r})\psi_{\lambda_2}(\mathbf{r}) + H.c. \tag{S132}$$

where $\Omega$ denotes the field strength. In the tight-binding model, it is

$$\hat{H}_1 = \sum_j \sum_{\lambda=g,e} \Omega \psi_{j,\lambda_1}^\dagger \psi_{j,\lambda_2} + H.c. \tag{S133}$$

It does not change under the transformation (S126). Then we make the following denotation,

$$\Omega = b_{xw} - ib_{yw} \,. \tag{S134}$$

In the momentum space, it gives

$$\hat{H}_1(\mathbf{k}) = b_{xw}\sigma_0 \otimes \sigma_x + b_{yw}\sigma_0 \otimes \sigma_y. \tag{S135}$$

which is identical to $\Delta_1$ of Hamiltonian (1) in the main text.

The perturbation term $\Delta_4$ of Hamiltonian (4) in the main text can be generated as follows. Likewise engineering the $A_w$ term, we introduce the auxiliary field as

$$\Delta_4 = \int d\mathbf{r} M_{\mathrm{aux}}(\mathbf{r})[\psi_{e_1}^\dagger(\mathbf{r})\psi_{g_1}(\mathbf{r}) + \psi_{e_2}^\dagger(\mathbf{r})\psi_{g_2}(\mathbf{r})] + H.c. \tag{S136}$$

with $M_{\mathrm{aux}}(\mathbf{r}) = -iM_{\mathrm{aux}}(\mathbf{r})[\cos(3k_Lx)\cos(k_Ly)\cos(k_Lz) - \cos(k_Lx)\cos(3k_Ly)\cos(k_Lz)]$. In the tight-binding model, the on-site term and the $z$-directional next-neighbor term vanish, leaving the $x$- and $y$-directional next-neighbor terms dominant but with an opposite sign. Therefore after making the transformation (S126), in the momentum space,

$$\Delta_4 = \delta[\cos(k_xd) - \cos(k_yd)] \tag{S137}$$

with $\int d\mathbf{r} i M_{\mathrm{aux}}(\mathbf{r}) W^*(\mathbf{r}) W(\mathbf{r} + \mathbf{e}_x) = -\int d\mathbf{r} i M_{\mathrm{aux}}(\mathbf{r}) W^*(\mathbf{r}) W(\mathbf{r} + \mathbf{e}_y) \equiv \delta$.

---



\end{document}